\documentclass[twocolappendix]{emulateapj}

\usepackage{natbib}
\usepackage{amsmath}
\usepackage{color}
\usepackage{ulem}
\definecolor{citecolor}{rgb}{0,0,0.5}
\usepackage[backref,breaklinks,colorlinks,citecolor=citecolor,linkcolor=magenta]{hyperref}
\usepackage[all]{hypcap}

\newcommand{\Athena}{{\it Athena}\ignorespaces}
\newcommand{\Hyperion}{{\it Hyperion}\ignorespaces}
\newcommand{\Orion}{{\it Orion}\ignorespaces}

\newcommand{\p}{\partial}

\newcommand\pc{\;{\rm pc}}
\newcommand\second{\;{\rm s}}
\newcommand\yr{\;{\rm yr}}
\newcommand\Myr{\;{\rm Myr}}
\newcommand\cm{\;{\rm cm}}
\newcommand\gram{\;{\rm g}}
\newcommand\erg{\;{\rm erg}}
\newcommand\kms{\;{\rm km}\,{\rm s}^{-1}}
\newcommand\Msun{\;M_{\odot}}
\newcommand\Lsun{{\;L_\odot}}
\newcommand\Kel{\;{\rm K}}
\newcommand\kB{{\;k_{\rm B}}}
\newcommand\eV{{\;{\rm eV}}}

\newcommand{\vecx}{\mathbf{x}}
\newcommand{\vecn}{\mathbf{n}}
\newcommand{\vecv}{\mathbf{v}}

\newcommand{\xn}{x_{\rm n}}
\newcommand{\xno}{x_{\rm 0}}
\newcommand{\xnp}{x_{\rm +}}
\newcommand{\xneq}{x_{\rm eq}}
\newcommand{\xneqo}{x_{\rm eq,0}}
\newcommand{\nH}{n_{\rm H}}
\newcommand{\nHI}{n_{\rm H^0}}
\newcommand{\nHII}{n_{\rm H^{+}}}
\newcommand{\nelec}{n_{\rm e}}
\newcommand{\HII}{\ion{H}{2}\ignorespaces}

\newcommand{\Erad}{\mathcal{E}}
\newcommand{\Frad}{\mathbf{F}}

\newcommand{\nuL}{\nu_{\rm L}}
\newcommand{\alphaB}{\alpha_{\rm B}}
\newcommand{\sigmaH}{\langle \sigma_{\rm H}\rangle}
\newcommand{\hnui}{h\nu_{\rm i}}

\newcommand{\SFE}{\varepsilon}
\newcommand{\myraylist}{{\tt my\_pp\_list}\ignorespaces}
\newcommand{\exitraylist}{{\tt exit\_pp\_list}\ignorespaces}

\begin{document}

\title{Modeling UV Radiation Feedback from Massive Stars: \\I.
  Implementation of Adaptive Ray Tracing Method and Tests} %

\shorttitle{Adaptive Ray Tracing in {\it Athena}} %

\shortauthors{Kim et al.} %








\author{Jeong-Gyu Kim\altaffilmark{1,2}, Woong-Tae
  Kim\altaffilmark{1}, Eve C.~Ostriker\altaffilmark{2}, \& M. Aaron
  Skinner\altaffilmark{2,3}}

\affil{$^1$Department of Physics \& Astronomy, Seoul National University, Seoul 08826, Republic of   Korea}
\affil{$^2$Department of Astrophysical Sciences, Princeton University, Princeton, NJ 08544, USA}
\affil{$^3$Lawrence Livermore National Laboratory, 7000 East Ave., Livermore,   CA 94550-9234, USA}
\email{jgkim@astro.snu.ac.kr, wkim@astro.snu.ac.kr, eco@astro.princeton.edu, skinner15@llnl.gov}

\begin{abstract}
  We present an implementation of an adaptive ray tracing (ART) module
  in the {\it Athena} hydrodynamics code that accurately and
  efficiently handles the radiative transfer involving multiple point
  sources on a three-dimensional Cartesian grid. We adopt a recently
  proposed parallel algorithm that uses non-blocking, asynchronous MPI
  communications to accelerate transport of rays across the
  computational domain. We validate our implementation through several
  standard test problems including the propagation of radiation in
  vacuum and the expansions of various types of \ion{H}{2} regions.
  Additionally, scaling tests show that the cost of a full ray trace
  per source remains comparable to that of the hydrodynamics update on
  up to $\sim 10^3$ processors. To demonstrate application of our ART
  implementation, we perform a simulation of star cluster formation in
  a marginally bound, turbulent cloud, finding that its star formation
  efficiency is $12\%$ when both radiation pressure forces and
  photoionization by UV radiation are treated. We directly compare the
  radiation forces computed from the ART scheme with that from the
  $M_1$ closure relation. Although the ART and $M_1$ schemes yield
  similar results on large scales, the latter is unable to resolve the
  radiation field accurately near individual point sources.
\end{abstract}


\section{INTRODUCTION}\label{s:intro}

Radiative feedback plays a vital role in regulating star formation on
various scales \citep[e.g.,][]{kru14}. On small scales, radiation from
(accreting) protostars raises the temperature of the surrounding gas,
affecting the evolution of individual accretion disks and suppressing
fragmentation of the dense cluster-forming gas into very low mass
objects \citep[e.g.,][]{kru06a,whi06,off09}. On intermediate scales,
ultraviolet (UV) radiation emitted by massive stars not only reduces
cold neutral gas available for star formation via photoionization and
photodissociation \citep{whi79,wil97,mat02,kru06b}, but also produces
thermal and radiation pressure that controls the dynamics of \HII\
regions by inducing expansion \citep{kru09,mur10,lop11,kim16}.
Ionizing radiation that escapes from star-forming regions also
produces diffuse ionized gas in the Milky Way and other galaxies
\citep[e.g.,][]{haf09} and contributes to re-ionization of the
intergalactic medium (IGM) in the Universe at high redshift
\citep[e.g.,][]{bar01}. On large scales in galaxies, far UV radiation
from young OB associations is the dominant heating source of the
diffuse neutral interstellar medium (ISM) via the photoelectric
effect on small dust grains \citep[e.g.,][]{wol03}. This controls the
thermal pressure that contributes to supporting the vertical
gravitational weight of the ISM in galactic disks, and thus represents
an important feedback loop that self-regulates star formation
\citep[e.g.,][]{ost10,kim13}. Therefore, it is essential to follow
effects of radiative feedback to address questions such as star
formation efficiencies and the mechanisms of cloud destruction for
giant molecular clouds (GMCs), as well as a wide variety of other
physical issues in the ISM and IGM.

Since star formation and feedback involve highly nonlinear processes,
radiation hydrodynamic (RHD) simulations have become an indispensable
tool in understanding the impact of radiation on star cluster
formation occurring in turbulent clouds. For example, several recent
studies relied on numerical simulations to investigate the effects of
photoionization \citep{wal12,dal12,dal13,how16,gav17} and radiation
pressure from dust-reprocessed infrared \citep{ski15} or stellar UV
\citep{ras16} radiation on cloud dispersal. \citet{dal14} and
\citet{dal17} studied the combined effects of ionizing radiation with
stellar winds. More recently, \citet{gee16}, \citet{gru16}, and
\citet{shi16} explored how the radiation feedback works together with
supernova explosions to destroy parent molecular clouds.

To treat radiation feedback properly, it is important to solve
radiative transfer (RT) equation accurately and efficiently. Despite
increasing demand for RHD simulations, RT still remains numerically
challenging for a number of reasons: high dimensionality, non-local
and multiscale behavior of interactions between radiation and matter,
and difficulty in choosing a suitable frame for evaluating radiation
and fluid variables, to name a few \citep{mih01,cas04}. A common
approach for the numerical solution of RT problems is to take the
angular moments of the transfer equation and to adopt a closure
relation to truncate the hierarchy of moments. As the time-dependent
moment equations can be written as hyperbolic conservation laws,
high-order Godunov methods are often adopted to solve the
time-dependent RT problem.

In astrophysics, the most widely-used moment method is the
flux-limited diffusion approximation, in which the radiation flux is
calculated by taking a local gradient of the radiation energy density
\citep{lev81,kru07a,gon15}, applying a limiter if the gradient is very
steep to prevent superluminal transport. RT solvers based on
flux-limited diffusion are suitable for describing radiation fields in
optically-thick fluids. However, they are of limited accuracy in
treating an optically-thin medium with a complex source distribution
as well as in casting shadows in the transition zones where the
optical depth varies significantly \citep[e.g.,][]{hay03,dav14}.

Another approach for RT is to use the $M_1$ closure relation that
assumes that the intensity is invariant under rotation about the
direction of radiation flux \citep[e.g.,][]{lev84,gon07,ski13,ros13}.
For a single point source, the $M_1$ closure model can correctly
describe the radiation field both in the optically thin and thick
regimes \citep[see][]{ski13}. When there are multiple sources
distributed widely, however, $M_1$ can fail in the optically-thin
regime. For example, when two beams going in different directions
interact with each other, they unphysically merge rather than crossing
\citep[e.g.,][]{fra12}, since the local flux is assumed to be
unidirectional. For diffuse radiation in systems with both optically
thin and thick regions, a more accurate closure relation can be
obtained from the formal solution of the RT equation using the
multidirectional method of short characteristics \citep{dav12}. For
point-like sources that have strong angular variations in emissivity,
however, artificial anisotropic structure can arise from inaccuracy in
interpolating intensity over neighboring cells in the short
characteristics method \citep{fin09}.

For problems in which the total emission is dominated by a small
number of point sources (and in which the reprocessed diffuse
radiation is negligible), one may directly integrate the RT equation
by calculating the column density (or, equivalently, the optical
depth) to every point in the simulation domain starting from the
sources. A method utilizing short characteristics calculates the
column densities from the sources by performing upwind interpolation
over cells \citep{mel06}. Alternatively, the long-characteristics
method computes the column densities to each zone by following rays
emitted by all sources until they reach the domain boundary
\citep{abe99,lim03}. This allows one to calculate the radiation field
more accurately over the entire domain at the expense of excessively
resolving the regions close to the sources when the number of ray
directions is large.

To alleviate the inefficiency of the traditional long characteristics
method, \citet{abe02} developed a novel, adaptive ray tracing (ART)
technique that has since been widely used to describe ionizing
radiation originating from massive stars, in various astronomical
contexts \citep[e.g.,][]{kru07b,bis09,wis11,bac15}. In the ART method,
rays are created at point sources and successively split as they are
traced outward based on the Hierarchical Equal Area isoLatitude
Pixelization (HEALPix) scheme \citep{gor05}. The salient feature of
this method is that via ray splitting, the angular resolution of
radiation adapts to the local hydrodynamical resolution, such that the
number of rays from each point source intersecting each grid zone
remains approximately constant. While this RT method is quite
efficient in achieving an accurate solution for the radiation field,
existing codes employing the ART method often suffer from poor
parallel performance. This is due to excessive overhead for
interprocessor communication as rays traverse the interfaces between
subdomains (see Section~\ref{s:parallel}), making some implementations
of the ART method essentially inapplicable to large-scale simulations
with multiple point sources.

Very recently, \citet{ros17} developed a hybrid RHD module for the
\Orion\ code by combining the ART method with the flux-limited
diffusion method. It makes use of the former to describe direct
radiation from point sources, while taking advantage of the latter to
treat diffuse radiation arising from gas and dust. Their parallel
algorithm for the ART uses completely non-blocking, asynchronous
communication, which greatly improves the parallel scaling of ray
tracing over previous implementations using a synchronous
communication algorithm. The scaling tests in \citet{ros17} showed
that the cost of the ART module in \Orion\ remains comparable to that
of hydrodynamics up to $\sim 10^3$ processors.

In this work, we describe implementation of an ART module for multiple
point sources in the grid-based magnetohydrodynamics (MHD) code
\Athena\ \citep{sto08} and present test results. Our implementation
closely follows the parallelization strategy proposed by
\citet{ros17} and includes a few new features that further improve
parallel performance. To model hydrogen ionization and recombination
processes, we adopt a simple and efficient explicit scheme based on
an analytic approximation. We first measure the scalability of our
implementation by performing weak and strong scaling tests. We then
apply the code to the standard test problems, namely, the expansion of
R-type and D-type ionization fronts as well as expansion of a dusty
\HII\ region driven by both thermal and radiation pressures. By
comparing the numerical results with analytic or semi-analytic
solutions, we demonstrate the accuracy of our ART implementation.

A key application for ART is to follow the detailed effects of
radiation produced by OB stars on their natal GMCs. To this end, we
have combined our ART module with other physics packages implemented
in \Athena\ and run numerical RHD simulations of star cluster formation
in turbulent, self-gravitating, unmagnetized clouds; these models are
similar to the RHD simulations of \citet{ski15,ras16}, but include
both ionizing and non-ionizing radiation computed using ART. In this
paper, we demonstrate practical application of our ART module via an
example simulation for a fiducial model. For comparison to the
radiation field computed using the $M_1$ closure relation in the
moment-based \Hyperion\ code \citep{ski13}, we have also run an RT model
with non-ionizing radiation only, using an identical set of sources
and density distribution. We find that the large-scale radiation
fields from the two RT methods are quite similar to each other, but
there is non-negligible difference at small scale near the point
sources, where resolution is inherently limited for pure moment
methods like that in \Hyperion.

The rest of this paper is organized as follows. In
Section~\ref{s:method}, we present the basic equations that we solve
and briefly review the algorithm of ART together with our
parallelization strategy. We also describe the subcycling method to
solve the equation for hydrogen ionization and recombination.
Section~\ref{s:test} presents the results of the scaling tests as well
as the tests of the expansion of \HII\ regions. In
Section~\ref{s:comp}, we describe the numerical setup for simulations
of star cluster formation in turbulent molecular clouds and compare
the radiation fields based on the ART method with those from the
moment method with the $M_1$ closure relation. Finally, we summarize
and discuss future applications of the code in
Section~\ref{s:summary}.

\section{NUMERICAL METHOD}\label{s:method}

Stellar UV radiation can alter the chemical, thermal, and dynamical
state of the ISM through various processes. In this paper, we consider
the two most basic processes: photoionization of hydrogen atoms and
direct radiation pressure applied to the gas/dust mixture. The
governing equations we adopt read:
\begin{equation}\label{e:cont}
  \dfrac{\p \rho}{\p t} + \nabla \cdot (\rho \vecv) = 0 \,,
\end{equation}
\begin{equation}\label{e:momentum}
  \dfrac{\p }{\p t} (\rho \vecv) + \nabla \cdot (\rho \vecv\vecv +
  P\mathbf{I}) = -\rho\nabla\Phi + \mathbf{f}_{\rm rad} \,,
\end{equation}
\begin{equation}\label{e:nHI}
  \dfrac{\p \nHI}{\p t} + \nabla \cdot (\nHI \vecv)
  = \mathcal{R} - \mathcal{I} \,,
\end{equation}
\begin{equation}\label{e:Poisson}
  \nabla\Phi = 4\pi G\rho\,,
\end{equation}
where $\rho = 1.4 m_{\rm H} \nH$ is the gas density with $\nH$
representing the total hydrogen number density, $\vecv$ is the
velocity, $P$ is the thermal pressure, $\Phi$ is the gravitational
potential, and $\mathbf{f}_{\rm rad}$ is the force per unit volume due
to absorption of stellar UV radiation. In Equation~\eqref{e:nHI},
$\nHI$ refers to the number density of neutral hydrogen, and
$\mathcal{I}$ and $\mathcal{R}$ denote the volumetric ionization and
recombination rates of hydrogen, respectively, whose functional forms
are given in Section \ref{s:source}. Numerical solution of Equations
(\ref{e:cont}) and (\ref{e:momentum}) uses the methods described by
\citet{sto08}, with the radiation force source term updated using an
operator splitting. The Poisson Equation (\ref{e:Poisson}) is solved
using fast Fourier transforms (FFTs) with vacuum boundary conditions
\citep{ski15}. At any location on the grid, the radiation flux and
energy density are required to compute the source terms
$\mathbf{f}_{\rm rad}$ and $\mathcal{I}$, respectively; these
radiation terms are computed via our implementation of ART.

In what follows, we describe how the radiation fields are computed in
the ART method in Section~\ref{s:ART}, while our method of
parallelization is presented in Section~\ref{s:parallel}. Our
algorithms for updating radiation source terms and advancing Equation
(\ref{e:nHI}) are described in Section~\ref{s:source}.

\subsection{Adaptive Ray Tracing}\label{s:ART}

Here we briefly review the RT problem for point sources in the
framework of ART. We refer the reader to \citet{wis11}, \cite{bac15},
and \citet{ros17} for more detailed descriptions.

We consider only the direct radiation field from multiple point
sources and ignore diffuse emission as well as scattered light. Under
the assumption that the light crossing time is much shorter than both
the sound crossing time and the time scale for the change in opacity,
streaming radiation reaches equilibrium with matter effectively
instantaneously. The radiation intensity should, therefore, satisfy
the time-independent transfer equation
\begin{equation}\label{e:RT}
  \vecn \cdot \nabla I = -\chi I\,,
\end{equation} where $I$ is the radiation intensity, $\chi =\rho \kappa$
is the (isotropic) extinction coefficient of matter per unit path
length for $\kappa$ the opacity, and $\vecn$ is a unit vector parallel
to the propagation direction of radiation. Both $I$ and $\chi$ depend
on the position ${\mathbf x}$, time $t$, and the frequency $\nu$,
which are suppressed for notational simplicity.

We solve Equation~\eqref{e:RT} along a set of rays that discretize the
directions in solid angle with respect to individual radiation
sources. For simplicity, we consider monochromatic radiation emergent
from a single point source located at $\vecx_{\rm src}$ with
luminosity $L$; an extension to polychromatic radiation from multiple
sources is straightforward by taking a summation over a discretized
set of frequencies and over rays from different sources. We inject
photon packets from $\vecx_{\rm src}$ and carry them radially outward
along the direction of any given ray by calculating the absorption
rates of the photon energy and momentum over the grid cells they pass
through. The direction of propagation $\vecn_{\rm ray}$ is determined
using the HEALPix scheme of \citet{gor05}, which divides the unit
sphere into $N_{\rm ray}(\ell)=12 \times 4^{\ell}$ equal-area pixels
at the level $\ell \ge 0$. Denoting the initial level by $\ell_0$,
each injected photon packet on a given ray carries luminosity
$L_{\rm ray}(r=0) = L/N_{\rm ray}(\ell_0)$, subtends a solid angle
$\Omega_{\rm ray}(\ell_0) = 4\pi/N_{\rm ray}(\ell_0)$ from the source,
and propagates along the ray
$\vecx = \vecx_{\rm src} + r \vecn_{\rm ray}$, where
$r = |\vecx - \vecx_{\rm src}|$ measures the distance from the source.

Integrating Equation~\eqref{e:RT} over the whole solid angle gives the
radiation energy equation
\begin{equation}\label{e:REE}
   \nabla \cdot\Frad = -c\chi\Erad\,,
\end{equation}
where $c$ is the speed of light, $\Erad$ is the radiation energy
density, and $\Frad$ is the radiation flux. With
$\Frad = \hat{\mathbf{r}}L e^{-\tau(r,\vecn)}/(4\pi r^2)$ and
$\Erad = |\Frad|/c$ for streaming radiation, Equation \eqref{e:REE}
can be written as
\begin{equation}\label{e:RT2}
  \dfrac{\p L_{\rm ray}}{\p r} = -\chi L_{\rm ray}\,,
\end{equation}
where
$L_{\rm ray}(r)= L e^{-\tau(r,\vecn_{\rm ray})}\Omega_{\rm
  ray}/(4\pi)$ is the discretized luminosity at distance $r$ in the
region subtended by $\Omega_{\rm ray}$. We compute the length of a
line segment $\Delta r$ between the two consecutive cell interfaces
intersected by a ray (see, e.g., \citealt{wis11,bac15}). The
corresponding cell optical depth is $\Delta \tau = \chi\Delta r$. The
absorption rates of the radiation energy and momentum by the material
along the path $\Delta r$ in a given cell are then
$\Delta L_{\rm ray} = L_{\rm ray,in} (1 - e^{-\Delta \tau})$ and
$\vecn_{\rm ray}\Delta L_{\rm ray}/c$, respectively, where
$L_{\rm ray,in}$ is the luminosity of the ray entering the cell. The
luminosity of a photon packet on a given ray is therefore reduced by
$\Delta L_{\rm ray}$ as it traverses the cell.

Utilizing the lab-frame equations of RHD \citep[e.g.,][]{mih01}, these
quantities are related to the volume-averaged radiation energy density
and flux in a given cell as
\begin{align}
  \Erad & = \dfrac{1}{\chi\Delta V}\sum_{\rm rays}\dfrac{\Delta L_{\rm
    ray}}{c}\,, \label{e:Erad}\\
  \Frad & = \dfrac{1}{\chi\Delta V}\sum_{\rm rays}\Delta L_{\rm
    ray} \vecn_{\rm ray}\,, \label{e:Frad}
\end{align}
where $\Delta V$ is the cell volume and the summation is taken over
all rays passing through the cell. Note that these expressions satisfy
the flux-limiting condition $|\Frad| \le c\Erad$ \citep{lev84} and
tend to the exact value in the limit of infinite angular resolution.
In updating Equation~\eqref{e:momentum} over time step $\Delta t$, an
amount $\Delta t \mathbf{f}_{\rm rad} = \Delta t \chi \Frad/c$ is
added to the gas momentum in each cell using Equation~\eqref{e:Frad}
for $\Frad$ (see Equation~\eqref{e:radforce} below; in practice these
operator-split increments to the gas momentum are applied after each
radiation subcycle). Since the photon momentum on a set of rays
traversing a cell is reduced by
$\sum_{\rm rays} \Delta L_{\rm ray} \vecn_{\rm ray}/c$ per unit time,
this update is manifestly conservative of momentum.

To ensure that each grid cell is sampled by at least $m_{\rm ray}$
rays, we split a parent ray into four child rays at one higher HEALPix
level if the solid angle $(\Delta x)^2/r^2$ (corresponding to the
maximum angle subtended by a given face of the current cell as seen
from the source) is smaller than $m_{\rm ray}\Omega_{\rm ray}$. With a
uniform grid spacing of $\Delta x$, this corresponds to the maximum
distance
$r_{\rm max}(\ell) = [3/(\pi m_{\rm ray})]^{1/2} 2^{\ell}\Delta x$
that rays can travel at level $\ell$. The new child rays are cast at
positions $\vecx_{\rm src} + r_{\rm max}(\ell)\vecn_{\rm ray}$. We
follow photon packets as they traverse rays, dividing them equally
when a parent ray splits into children. A ray is terminated and its
photon packet is destroyed either where a ray exits the computational
domain or where the photon packet is completely absorbed. The latter
occurs when the total optical depth $\tau(r,\vecn_{\rm ray})$ from the
source is larger than a specified value $\tau_{\rm max}$, which we set
to 7 as a default value. In order to alleviate the geometrical
artifacts arising from the use of the crossing length rather than the
ray-cell volume intersection, we randomly rotate the injection
directions of the rays at every time step \citep{kru07b}.

\begin{figure*}[!t]
  \epsscale{1.0}\plotone{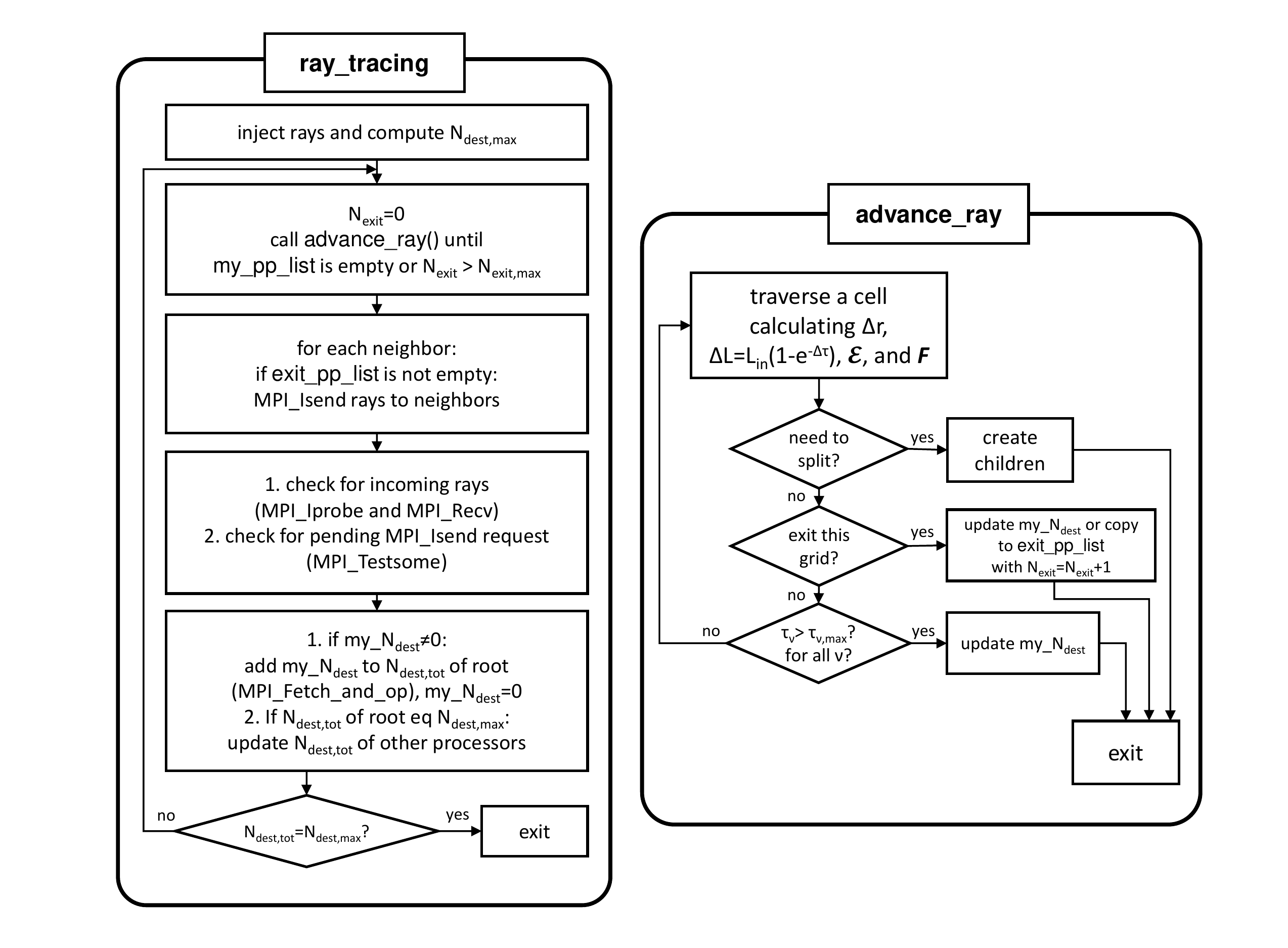}
  \caption{Flowchart of the ART algorithm: {\tt ray\_tracing} performs
    one ray tracing throughout the global computational domain while {\tt
      advance\_ray} traces one ray within a local subdomain (``grid'' in
    \Athena).}\label{f:cflow}
\end{figure*}

\subsection{Parallelization}\label{s:parallel}

Although conceptually simple and easy to implement, the major drawback
of the ART method so far has been its poor parallel performance. There
are two chief impediments to achieving scalable performance on a
distributed memory platform: load imbalance and communication
overhead. First, the amount of work needed per processor for the ART
is roughly proportional to the number of the ray-cell crossings. Since
this depends not only on the spatial distribution of sources but also
on the opacity of the material, processors most likely have unequal
workload depending on the problem geometry. This imposes an inherent
limit to scalability of ART with static domain decomposition. One may
alleviate this issue by dynamically adjusting the local domain sizes
to evenly distribute the workload.

The second problem is more serious and common to long-characteristic
methods implemented with domain decomposition, in which the data from
different subdomains are local to individual processors. For rays
traveling across multiple subdomains, processors need to share
information such as the propagation directions of the rays, optical
depth, etc., in order to integrate Equation~\eqref{e:RT2} along the
characteristics. Because a ray may be terminated by the
$\tau > \tau_{\rm max}$ condition before traversing a subdomain that
lies along the ${\bf n}_{\rm ray}$ direction, it cannot be determined
in advance when, from where, or how many rays will enter a particular
subdomain. The size and pattern of the data that need to be
communicated among processors are therefore highly irregular. This
poses a major challenge to the parallelization of the ART code.
Indeed, the communication overhead has been the dominant performance
bottleneck in the existing long-characteristic methods
\citep[e.g.,][]{rij06,wis11}.

\citet{ros17} presented an efficient parallel algorithm for the ART
module implemented in \Orion, an adaptive mesh refinement (AMR) code.
The key idea in their algorithm is to make use of the non-blocking
message-passing operations ({\tt MPI\_Isend}, {\tt MPI\_Iprobe}, and
{\tt MPI\_Test}) provided in the MPI-3
standard\footnote{\href{http://mpi-forum.org/docs/mpi-3.1}{http://mpi-forum.org/docs/mpi-3.1}}
for communication of ray information between processors. This allows
processors to carry on their work (local transport of photon packets along
rays) while communication is pending (see also \citealt{bac15}). Thus,
the transport of rays within and between subdomains can be executed
asynchronously, greatly reducing the idle time spent waiting for other
processors to finish their work.

Another hallmark of the \citet{ros17} algorithm is the use of a global
``destroy'' counter $N_{\rm dest}$, which keeps track of how many rays
have been terminated in the whole computational domain. Whenever a ray
at the HEALPix level $\ell$ is terminated, we increment the local
destroy counter by $N_{\rm dest,ray} = 4^{\ell_{\rm max} - \ell}$,
where $\ell_{\rm max}$ is the maximum level allowed. The global
destroy counter is then a simple sum of the local destroy counters,
and it stays synchronized across the processors by non-blocking
communications. When all rays are terminated, we have
$N_{\rm dest}=N_{\rm dest,max} = N_{\rm src} \times 12 \times
4^{\ell_{\rm max}}$, where $N_{\rm src}$ is the number of point
sources in the entire domain. Updating the destroy counter as a global
shared variable allows processors to determine when to exit the
work-communication cycle without relying on synchronous, blocking
communication.

We have implemented ART in \Athena\ following the parallelization
approach of \citet{ros17}, with some additional modifications to
improve parallel performance. The schematic overview of the ray
tracing algorithm is shown in \autoref{f:cflow}. Our ART algorithm
runs the following steps:

\begin{description}
\item[Initialization] (Executed only once at the start of the
  simulation) Find neighbor grids\footnote{In \Athena\ without mesh
    refinement, the computational domain is divided into a set of
    rectangular ``grids'', each of which is owned by a single
    processor. Thus, ``grid'' is equivalent to ``subdomain'' for
    \Athena.} that share faces, edges, or corners, and allocate memory
  for arrays \myraylist\ and \exitraylist, which will store local and
  outgoing ray (or photon packet) information, respectively. Each
  processor has one \myraylist\ and $n_{\rm ngbr}$ \exitraylist, where
  $n_{\rm ngbr}$ is the number of the neighbor grids.
\item[Step 1] If any sources exist within my grid, create
  an initial set of directions $\vecn_{\rm ray}$
  and corresponding $L_{\rm ray}$ values
  and store the photon packet information
  in \myraylist. Compute $N_{\rm dest,max}$ from  $N_{\rm src}$
  summed over all grids.
\item[Step 2] Transport photon packets along their respective
  $\vecn_{\rm ray}$ directions within the local domain until either
  the total number of photon packets that need to be passed to
  neighbor grids exceeds $n_{\rm exit,max}$ or no local ray is left in
  \myraylist. Each ray is followed until it (1) needs to split, (2)
  reaches the grid boundaries, or (3) is terminated with
  $\tau \ge \tau_{\rm max}$. Stack information on photon packets for
  rays leaving the local grid in the particular \exitraylist\ for each
  neighbor. Calculate radiation energy density and flux for cells
  through which rays pass, as described in Section~\ref{s:ART}.
\item[Step 3] Loop over the list of the neighbors and send photon
  packet information in \exitraylist\ to the neighbor grids ({\tt
    MPI\_Isend}), checking if the previous operation has completed
  ({\tt MPI\_Testsome}).
\item[Step 4] Check for incoming messages from neighbors ({\tt
MPI\_Iprobe}). If there are any, execute a blocking receive for each
of them ({\tt MPI\_Recv}) and copy the received data to \myraylist.
\item[Step 5] If \myraylist\ is empty, add $N_{\rm dest}$ to the
global destroy counter $N_{\rm dest,tot}$ in the root processor ({\tt
MPI\_Fetch\_and\_op}) and reset $N_{\rm dest}$ to 0.
\item[Step 5-1] (Root processor only) If $N_{\rm dest,tot}=N_{\rm
dest,max}$, update $N_{\rm dest,tot}$ to $N_{\rm dest,max}$ in the
other processors ({\tt MPI\_Fetch\_and\_op}).
\item[Step 6] Go back to Step 2 if $N_{\rm dest,tot} \ne N_{\rm
dest,max}$. Exit the ray tracing when $N_{\rm dest,tot} = N_{\rm
dest,max}$.
\end{description}

Our implementation of the ART method has two notable differences
compared to that of \citet{ros17}. First, in \citet{ros17}, photon
packets in a given grid are communicated to neighbors only after all
rays have been traced to the grid boundaries. Therefore, processors
without a source have to wait, repeatedly checking for rays coming
from neighbors, until the intervening grids toward the sources
successively finish their work. In our implementation, instead of
waiting until all local ray-tracing is complete, communications
between neighbors are initiated as soon as the number of rays traced
to the subdomain boundary exceeds a certain prescribed number
$n_{\rm exit,max}$, reducing the idle time spent by downstream
processors (see also \citealt{bac15}). Although it is ideal to
communicate a single photon packet right after the ray-tracing reaches
the subdomain boundary to keep the downstream processors busy whenever
possible, there is an optimal granularity, i.e., ratio of computation
to communication, or an optimal value of $n_{\rm exit,max}$ that gives
the best performance due to the finite overhead required for function
calls and synchronization. The choice of $n_{\rm exit,max}$ certainly
depends on the source distribution, the domain decomposition, and
machine specifications. In the case of a single point source at the
box center, we find $n_{\rm exit,max}=100$ gives the best performance,
and we adopt this value for the rest of this paper.

Second, unlike in \citet{ros17}, we perform atomic (that is,
uninterruptible) memory updates\footnote{Atomic operations are
  guaranteed to complete without interference from other operations,
  ensuring correct results even when multiple processors try to access
  $N_{\rm dest,tot}$ simultaneously.} of the destroy counter utilizing
the {\tt MPI\_Fetch\_and\_op} function, which is one of the
Remote-Memory-Access (RMA) operations in the MPI library. In the RMA
operations, also called one-sided communications, processors are
allowed to access a remote target memory without involving the
explicit intervention of the remote processor. The RMA operations have
comparatively low overheads since one processor specifies all
communication parameters. We find that the use of the {\tt
  MPI\_Fetch\_and\_op} function helps to reduce substantially the
overall cost of ray tracing when a large number of processors
($N_{\rm core} \geq 256$) are used.

In addition to these differences, we find that the following
implementation details can further improve parallel efficiency.

\begin{itemize}
\item We use arrays for storing a collection of local and
  outgoing/incoming rays, to enable copying data to/from the
  communication buffer in one chunk. We dynamically allocate memory
  for arrays, which doubles in size if the old array becomes full.
\item As pointed out by \citet{bac15}, we store only necessary
  information required for ray tracing, splitting, etc.\ in the ray
  data structure to minimize the amount of data transferred between
  processors. These include the HEALPix level $\ell$, HEALPix pixel
  number, luminosity $L_{\rm ray}$, source position
  $\mathbf{x}_{\rm src}$, ray direction $\mathbf{n}_{\rm ray}$,
  distance from the source, and the (integer) indices of the current
  cell in a grid.
\item We treat \myraylist\ as a stack and access its elements in a
  Last-In-First-Out manner. For example, if a parent ray splits, child
  rays are stacked immediately on top of one another or on a sibling
  of the parent. This implies that we recursively follow a ray and its
  child until the grid boundary is reached (see also \citealt{bac15}).
\item In Step 1, photon packets are stacked in such a way that every
  12 contiguous blocks of elements in \myraylist\ have ray directions
  that belong to the distinct 12 pixels at the base HEALPix level. The
  pixel numbers (in the nested numbering scheme) are arranged in 12
  hierarchical tree structures, corresponding to the 12 base-level
  pixels. With $\ell_0=4$, for example, the HEALPix numbers of
  injected rays can be ordered as
  $0,4^4,2\times 4^4,\cdots, 11\times 4^4, 1,
  1+4^4,1+2\times4^4,\cdots$.
\end{itemize}

All of the above help to distribute workload to downstream processors
as fast as possible and at a similar rate in all directions.

\subsection{Update of Radiation Source Terms}\label{s:source}

We make a number of simplifying assumptions in modeling the
photochemistry of hydrogen and its thermodynamic state. First, we
adopt the on-the-spot approximation in which every diffuse
Lyman-continuum photon resulting from a recombination to the ground
state is locally reabsorbed. We neglect the diffuse ionizing radiation
that can be important in the evolution of radiatively-driven collapse
and instabilities of ionization fronts \citep{haw12}. Second, we
do not consider collisional ionizations, which are negligible compared
to photoionizations in the temperature range of our interest
($T\lesssim 10^4\rm\,K$). Third, we do not solve an energy equation
that accounts for radiative heating and cooling. Instead, we simply
assign the gas temperature according to
\begin{equation}\label{e:Tiso}
  T = T_{\rm ion} - \left(\dfrac{\xn}{2 - \xn}\right) (T_{\rm ion} -
  T_{\rm neu})\,,
\end{equation}
where $\xn \equiv \nHI/\nH$ is the neutral gas fraction, and
$T_{\rm ion}$ and $T_{\rm neu}$ are the prescribed temperatures of the
fully ionized ($\xn=0$) and purely neutral ($\xn=1$) states,
respectively \citep[e.g.,][]{hen05,kim16}.\footnote{In reality, the
  gas temperature profile exhibits a peak immediate behind an
  ionization front due to spectral hardening. However, \citet{lef94}
  found that the detailed functional form of $T(\xn)$ does not
  significantly affect the dynamics of H\,{\tiny II} regions.} For our
tests presented in this work, we adopt the constant temperatures
$T_{\rm ion} = 8000 \Kel$ and $T_{\rm neu} = 20 \Kel$, the latter of
which falls within the temperature range inferred from the line ratios
of low-$J$ rotational transitions of CO molecules (e.g.,
\citealt{yod10}). The gas thermal pressure is then set by
$P = [1.1 + (1 - \xn)]\nH \kB T$, accounting for a 10\% of helium
content. The two-temperature isothermal equation of state has been
adopted by numerous numerical studies of \HII\ regions and is valid as
long as the time scale for approaching thermal equilibrium is short
compared to the dynamical time scale
\citep[e.g.,][]{lef94,wil02,gri09,mac14,ste17}. Although idealized,
Equation \eqref{e:Tiso} is a simple and practical approach to
following the pressure-driven dynamical expansion and internal
structure of \HII\ regions.

Equation~\eqref{e:nHI} describes temporal changes of the neutral
hydrogen fraction, with the recombination and ionization rates given
by
\begin{equation}
  \mathcal{R} = \alphaB \nelec \nHII\,,
\end{equation}
and
\begin{equation}\label{e:photrate}
  \mathcal{I} = \nHI
  \int_{\nuL}^{\infty}\dfrac{c\Erad_{\nu}}{h\nu}\sigma_{\rm
    H}(\nu) d\nu\,,
\end{equation}
respectively, where $\Erad_{\nu}$ is the radiation energy density per
unit frequency, $\alphaB = 3.03 \times 10^{-13}$
$ (T/8000 \Kel)^{-0.7} \cm^3\second^{-1}$ is the case B recombination
coefficient \citep{ost89,kru07b}, $\nHII = \nelec=\nH(1-\xn)$ is the
number density of protons and electrons, $\nuL$ is the Lyman limit
corresponding to $h\nuL=13.6 \eV$, and $\sigma_{\rm H}(\nu)$ is the
photoionization cross section. In practice,
Equation~\eqref{e:photrate} is evaluated as a discrete summation over
a finite number of frequency bins. In this work, we use one frequency
bin for ionizing radiation, denoted by the subscript ``${\rm i}$''.
Equation~\eqref{e:photrate} then becomes
\begin{align}
  \mathcal{I} & = \nHI \dfrac{c \Erad_{\rm i}}{\hnui} \sigmaH \\
              & = \dfrac{1}{\Delta V} \dfrac{\nHI \sigmaH}{\chi_{\rm i}} \sum_{\rm rays} \dfrac{\Delta L_{\rm
                ray,i}}{\hnui}\,,\label{e:photrate2}
\end{align}
where $\Erad_{\rm i}=\int_{\nuL}^{\infty} \Erad_{\nu} d\nu$ is the energy
density of ionizing radiation, as evaluated using Equation~\eqref{e:Erad},
$\hnui$ is the mean energy of the ionizing photons, and
\begin{align}\label{e:sigmaH}
  \sigmaH =\int_{\nuL}^{\infty} \dfrac{\Erad_{\nu}/h\nu}{\Erad_{\rm i}/\hnui}
  \sigma_{\rm H}(\nu) d\nu\,,
\end{align}
is the frequency-averaged effective cross section. The actual value of
$\sigmaH$ depends on the spectral shape of the incident radiation
$\Erad_{\nu}$, which in turn depends both on the optical depth from
the source and the spectral types of ionizing stars. For simplicity,
we take the constant values $\hnui=18 \eV$ and
$\sigmaH=6.3\times 10^{-18}\cm^2$ in the present work.

We solve Equation~\eqref{e:nHI} in two steps: hydrodynamic update and
source update by treating $\mathcal{R}$ and $\mathcal{I}$ as source
terms. The source update requires solving
\begin{equation}\label{e:xn}
  \dfrac{d\xn}{dt}=\frac{\mathcal{R} - \mathcal{I}}{\nH} = \alphaB\nelec(1 - \xn)- \xn\Gamma\,,
\end{equation}
where $\Gamma \equiv \mathcal{I}/\nHI$. Since the time scale for
changes of $\xn$ is almost always smaller than the hydrodynamic time
step, we update $\xn$ explicitly using subcycling, as explained below.

Assuming that $\alphaB$ and $\Gamma$ are constant during a substep and
using $\nelec=\nH(1-\xn)$, the right-hand side of
Equation~\eqref{e:xn} is quadratic in $\xn$. \citet{alt13} showed that
this has an analytic solution\footnote{These expressions can be
  generalized to include the collisional ionizations as well as the
  contribution to $\nelec$ from heavy elements. See Appendix C3 in
  \citet{alt13}.}
\begin{equation}\label{e:xn2}
  \xn(t) = \xneq + \dfrac{(\xnp - \xneq)(\xno - \xneq)K}{(\xnp - \xno)
    + (\xno - \xneq)K}\,,
\end{equation}
where $x_0 = \xn(t_0)$,
$K = \exp\left[-(\xnp - \xneq)(t-t_0)\alphaB\nH)\right]$,
$\xnp = \xneq^{-1}$, and
\begin{equation}
  \xneq = \dfrac{2\alphaB\nH}{\Gamma + 2\alphaB\nH + \sqrt{\Gamma^2 +
      4\alphaB\nH\Gamma}}
\end{equation}
is the equilibrium neutral fraction.
When $\Gamma=0$, Equation~\eqref{e:xn} has a solution
\begin{equation}
  \xn(t) = \dfrac{\xno + (1 - \xno)\alphaB\nH t}{1 + (1 -
    \xno)\alphaB\nH t}\,,
\end{equation}
which should be applied to cells completely shielded from ionizing
radiation. While Equation~\eqref{e:xn2} is exact, it is not
computationally robust when the denominator is close to zero, possibly
resulting in inaccurate $\xn$ due to amplified roundoff errors.

Alternatively, if we (incorrectly) treat $\nelec$ as being constant,
Equation \eqref{e:xn} yields a solution
\begin{equation}\label{e:xn3}
  \xn(t) = \xneqo + (\xno - \xneqo)e^{-(t-t_0)/t_{\rm i\textnormal{-}r}}\,,
\end{equation}
where $t_{\rm i\textnormal{-}r} = (\Gamma + \alphaB\nelec)^{-1}$ is the
ionization-recombination time and
$\xneqo = \alphaB\nelec/(\Gamma + \alphaB\nelec)$ is the equilibrium
neutral fraction \citep[e.g.,][]{sch87,mel06}. \citet{mel06} adopted
Equation~\eqref{e:xn3} to implicitly update the time-averaged
ionization fraction in their C$^2$-ray method. In the Appendix, we present
the test results of Equation~\eqref{e:xn3} on the temporal changes of
$\xn$ in a single cell exposed to a fixed radiation field. It turns
out that although Equation~\eqref{e:xn3} is based on the incorrect
assumption of constant $\nelec$, in practice it gives almost identical
results to those with Equation~\eqref{e:xn}. In addition,
Equation~\eqref{e:xn3} is robust and guarantees that $\xn$ always lies
between $\xno$ and $\xneqo$. We also find (see Appendix)
that it is more accurate
than methods based on the backward-difference formula often used in
the literature \citep[e.g.,][]{ann97}. We thus use
Equation~\eqref{e:xn3} to calculate $\xn$ in our implementation of the
subcycle.

We determine the size of substeps as
\begin{align}\label{e:dtss}
  \Delta t_{\rm ss} & = C\times {\rm min} \left( \dfrac{\nH}{\lvert
                      \mathcal{I} - \mathcal{R}\rvert} \right)\,,
\end{align}
with a constant coefficient $C$. Taking $C=0.1$ restricts the change
of $\xn$ per substep to below $0.1$ (see also \citealt{bac15}). The
minimum value of $\Delta t_{\rm ss}$ is usually from the cells in
transition layers where $0.1 \lesssim \xn \lesssim 1$ and
$\xneq \approx 0$.

\begin{figure*}[!t]
  \epsscale{1.0}\plotone{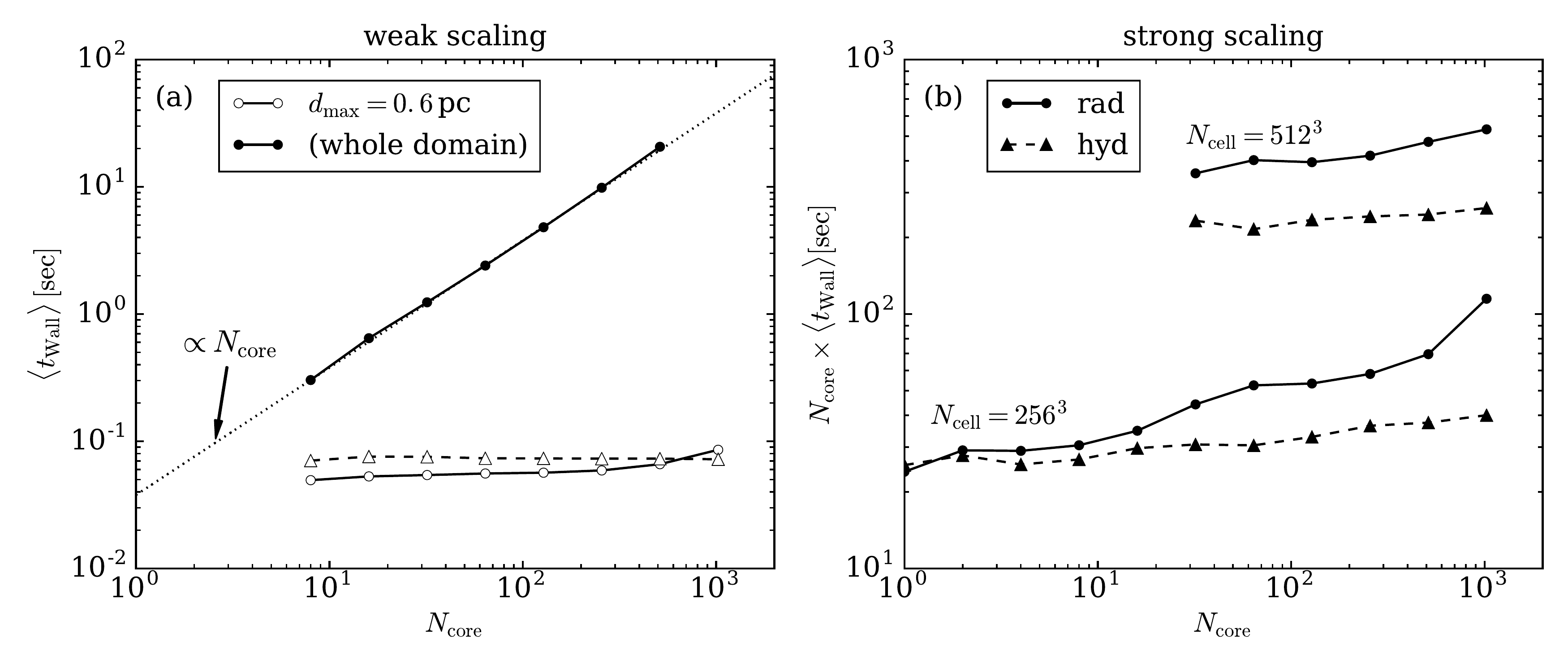}
  \caption{(a) Wall clock time {\it vs.} $N_{\rm core}$ in the weak
    scaling test for the ART (circles) and hydrodynamics solver
    (triangles) in which every $32^3$ grid belonging to a given
    processor has a single point source at the center. The open
    circles correspond the case in which each ray is terminated
    $0.6 \pc$ from its originating source, while filled circles denote
    the case in which all rays are extended until they hit the
    simulation domain boundary. (b) Wall clock time multiplied by
    $N_{\rm core}$ {\it vs.} $N_{\rm core}$ in the strong scaling
    test. A single point source is placed at the center of the box
    with 256 or 512 cells per side.}\label{f:scaling}
\end{figure*}

Our overall computation procedure is as follows.
We first evolve the hyperbolic terms in
Equations~\eqref{e:cont}--\eqref{e:nHI} for a full hydrodynamic time
step $\Delta t$ using the existing Godunov-type scheme in the \Athena\
code. The total gas density $\nH$ and the neutral gas density $\nHI$
are then available as inputs to the ART module for radiation (and
ionization/recombination) subcycles. Next, we perform the ART for each
radiation subcycle to calculate $\Erad$ and $\Gamma$ and determine
$\Delta t_{\rm ss}$. We then update the neutral fraction by using
Equation~\eqref{e:xn3}; given an updated $\xn$, the electron density
is updated to $n_e=\nH(1-\xn)$. At each subcycle over
$\Delta t_{\rm ss}$, we explicitly calculate the radiation force
\begin{equation}\label{e:radforce}
   \mathbf{f}_{\rm rad} = \frac{1}{c} \sum_{\nu_j}
   \chi_{\nu_j}\Frad_{\nu_j}\,,
\end{equation}
using Equation (\ref{e:Frad}) for each frequency bin and add
$\mathbf{f}_{\rm rad}\Delta t_{\rm ss}$ to the gas momentum density if
radiation pressure is switched on. The total opacity in the ionizing
and non-ionizing bins is calculated as
\begin{equation}
  \chi_{\rm i} = \nHI \sigmaH + \nH \sigma_{\rm d}\,,
\end{equation}
and
\begin{equation}
  \chi_{\rm n} = \nH \sigma_{\rm d}\,,
\end{equation}
respectively, where we use the constant-attenuation cross section per
hydrogen atom
$\sigma_{\rm d} = 1.17 \times 10^{-21} \cm^2\,\rm{H}^{-1}$
\citep{dra11}.

\section{RESULTS OF CODE TESTS}\label{s:test}

We now present the results of various tests intended to verify the
performance and accuracy of our implementation of the ART, including
its ability to simulate the dynamics of \HII\ regions. Unless otherwise
noted, we adopt $m_{\rm ray}=4$ and $\ell_0=4$ as fiducial values for
the angular resolution and the initial HEALPix level, respectively
\citep{bac15,ros17}.

\subsection{Scaling Tests}

To measure the parallel performance of the ART and its cost relative
to that of the hydrodynamic solver, we conduct weak and strong scaling tests
similar to the ones presented in \citet{ros17}. The tests are run on
16-core Intel Sandy Bridge nodes in the Tiger cluster at Princeton
University.

In the weak scaling test, the whole computational domain is subdivided
into $N_{\rm core}$ identical grids, each having $32^3$ cells over a
$1 \pc^3$ volume with a source at each grid center. Each grid is
assigned to a processor, and we time the execution of the ART as well
as the hydrodynamics solver, varying the number of processors
$N_{\rm core}$. \autoref{f:scaling}(a) shows the wall clock time
$\langle t_{\rm Wall} \rangle$, averaged over 10 cycles, taken for a
single ART (circles) and hydrodynamics (triangles) update. We consider
problem sizes between $N_{\rm core} = 2^3$ and $2^{10}$. The open
circles correspond to the case in which rays are terminated at
$d_{\rm max} = 0.6 \pc$, such that communication is only with
immediately adjacent neighbors. Overall, the wallclock time is nearly
flat with varying $N_{\rm core}$, indicating excellent parallel
efficiency (a horizontal line would represent perfect weak scaling for
this case). Quantitatively, the runtime to perform the ray tracing is
$0.050 \second$ when $N_{\rm core} = 8$, slightly shorter than the
hydrodynamics update, and increases to $0.086 \second$ when
$N_{\rm core} = 2^{10}$, with parallel efficiency of $58\%$. For the
weak scaling test, we also consider the case where rays are followed
until they exit the simulation domain, which is plotted as filled
circles. This is in good agreement with the prediction of the perfect
scaling for a processor workload that increases with the number of
sources, and hence, rays per cell
$\propto N_{\rm source} \propto N_{\rm core}$, with
$\langle t_{\rm Wall} \rangle \propto N_{\rm core}$ shown as a dotted
line, indicating that the processors are busy most of the time.

In the strong scaling test, the total problem size remains fixed, with
the domain decomposed into varying number of grids. We consider a
domain with $256^3$ or $512^3$ cells and add a single point source at
the domain center. We do not restrict the distance that rays can
travel from the source: rays extend to the domain boundaries.
\autoref{f:scaling}(b) plots the wall clock time to complete one
ray tracing (solid) and hydrodynamics update (dashed) multiplied by
$N_{\rm core}$. A horizontal line would represent perfect strong
scaling for this case. For the domain with $256^3$ cells, the run time
increases by a factor of $4.74$ as $N_{\rm core}$ varies from 1 to
1024, corresponding to a scaling efficiency of 21\%. For the domain
with $512^3$ cells, we start from $N_{\rm core}=32$, since the problem
becomes memory bound, and obtain a relative parallel efficiency of
$67\%$ on 1024 cores. It is remarkable that the cost of ray tracing in
our implementation is comparable to that of hydrodynamic updates even
though rays pass through multiple subdomains to reach the domain
boundaries. This indicates that our communication approach is
successfully distributing work throughout the domain, and in
particular, that ``downstream'' processors do not suffer from being
idle.

Compared to the strong scaling presented by \citet{ros17} (their
Figure 8), our result for the $256^3$ box shows a performance
improvement by a factor of more than 10. Differences in machine
specifications and hierarchical grid structures may contribute to
these differences in scaling results. For instance, the \Orion\ AMR
code used by \citet{ros17} employs a patch-based AMR method in which
the domain is decomposed into a set of grids of uniform cell spacing,
and multiple grids may be assigned to a single core. The same
patch-based method is applied even without mesh refinement. Finding
neighbors on the patch-based mesh is non-trivial and requires looping
over local grids until the next grid is found. This is in contrast to
the domain decomposition method adopted by \Athena, in which a single
core covers only a single grid. Therefore, the patch-based scheme
requires the additional cost of finding neighbor grids when a ray
needs to be passed between grids even when those grids reside on a
single core. This most likely accounts for some of the differences in
the scaling results. Further optimizations described in Section
\ref{s:parallel} may also contribute to the performance improvement.

\subsection{Radiation in a Vacuum}\label{s:vacuum}

We now assess the accuracy of our implementation of the ART method for
recovering the inverse-square law of radiation energy density around a point
source in vacuum. With a limited angular resolution, the ART method is
unable to achieve perfect spherical symmetry under Cartesian geometry.
We study the effects of the angular resolution parameter $m_{\rm ray}$
as well as the ray rotation on the accuracy of the calculated radiation
energy density.

\begin{figure}[!t]
  \epsscale{1.1}\plotone{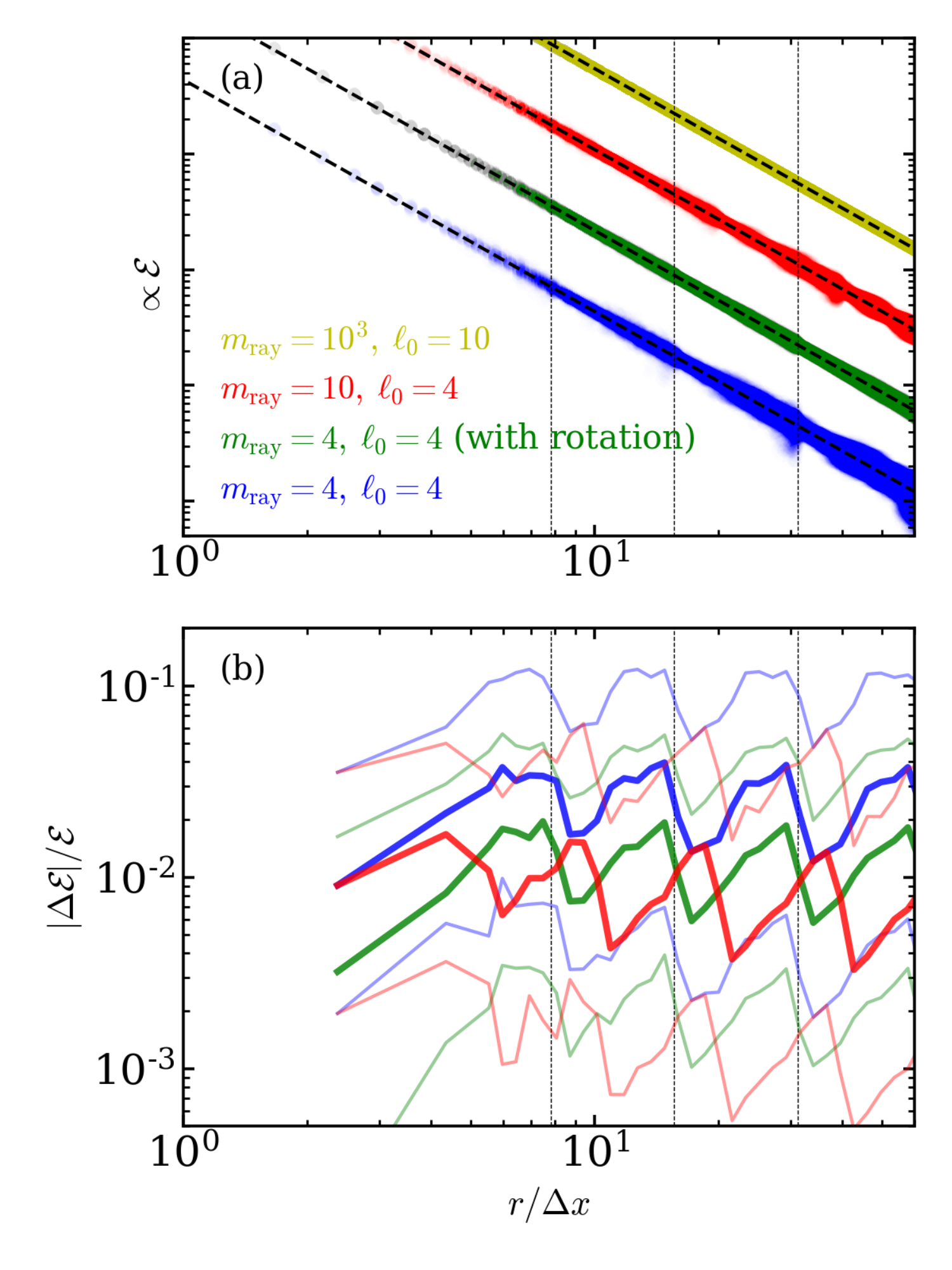}
  \caption{Test of the ART on the radiation field around a single
    point source in vacuum. (a) Radiation energy density {\it vs.} the
    distance from the source (normalized by the grid spacing
    $\Delta x$) with $(m_{\rm ray},\ell_0)=(4,4)$ (blue and green),
    $(m_{\rm ray},\ell_0)=(10,4)$ (red), and
    $(m_{\rm ray},\ell_0)=(10^3,10)$ (yellow). The data for each case
    are shifted along the ordinate for clear comparison. The green
    dots show the averages of ten ARTs, each with different ray
    orientation. The black dashed lines denote the analytic solution
    $\Erad \propto r^{-2}$. (b) The $10^{\rm th}$, $50^{\rm th}$
    (median; heavy line), and $90^{\rm th}$ percentile values of the
    relative error in each radial bin. The vertical dashed lines
    indicate the distance at which rays split for
    $(m_{\rm ray},\ell_0)=(4,4)$.}\label{f:vacuum}
\end{figure}

As a computational domain, we consider a cubic box with a side length
of $2\pc$ and place a point source with luminosity $L=1\Lsun$ at the
center. The box is discretized into $128^3$ cells.
\autoref{f:vacuum}(a) shows the distributions of the
volume-averaged radiation energy density $\Erad$ for
$(m_{\rm ray},\ell_0)=(4,4)$ (blue), $(m_{\rm ray},\ell_0)=(10,4)$
(red), and $(m_{\rm ray},\ell_0)=(10^3,10)$ (yellow) against the
normalized distance from the source. Each distribution is shifted by a
constant factor along the ordinate for clear comparison. The
deviations of $\Erad$ relative to the cell-centered value
$L/(4\pi r^2 c)$, plotted as the black dashed line, is mostly on the
order of a few percent for $m_{\rm ray}=4$. We also run 10 different
instances of the ray-trace with $m_{\rm ray}=4$, randomly varying the
ray orientation. The mean values of the resulting $\Erad$ are plotted
as green symbols, demonstrating that the errors introduced by
geometrical artifacts can be reduced by rotating the directions of ray
injection \citep{kru07b}.

\autoref{f:vacuum}(b) plots as solid lines the 10th, 50th (median),
and 90th percentiles of the relative errors $|\Delta \Erad|/\Erad$
within spherical shells centered at the source, after taking the case
with $(m_{\rm ray},\ell_0)=(10^3,10)$ as the reference solution. The
median value of the relative error is $1$--$4\%$ for
$(m_{\rm ray},\ell_0)=(4,4)$. The case with ray rotation achieves a
median accuracy similar to that with $m_{\rm ray}=10$. The sawtooth
patterns in the relative errors reflect the radial variation of the
angular resolution: the resolution becomes gradually worse with
increasing $r$ at a given HEALPix level and suddenly increases at the
ray-splitting radii $r_{\rm max}(\ell)$, plotted as the vertical
dashed lines for $(m_{\rm ray},\ell_0)=(4,4)$. Varying $m_{\rm ray}$
from 2 to 256, we find that the median value of the relative errors
can be fitted as $2.0 (m_{\rm ray}/4)^{-1.32}\%$, which is steeper
than $\propto m_{\rm ray}^{-0.6}$ obtained by \citet{wis11}.

\subsection{Expansion of \HII\ Regions}

Next we perform three tests of the expansion of an \HII\ region
embedded in a uniform medium. These are classical problems in
astrophysics with well-known analytic solutions, and have thus been
the standard tests for RHD codes incorporating the
effects of ionization and recombination
\citep[e.g.,][]{mel06,kru07b,wis11,bac15,bis15}. In this section, we
adopt $T_{\rm neu} = 100 \Kel$ and $T_{\rm ion} = 8000 \Kel$.

\subsubsection{R-type Ionization Front}

We consider an R-type ionization front created by a central source in
a uniform static medium with hydrogen number density $\nH$. At $t=0$,
the source starts to emit ionizing photons at a constant rate of
$Q_{\rm i}$. The expansion of the ionization front is very rapid at
early time, without inducing significant gas motions. If there is no
dust extinction and the recombination coefficient is taken as constant,
the ionization front expands as
\begin{equation}\label{e:Rtype}
  r_{\rm IF} = R_{\rm St,0} (1 - e^{-t/t_{\rm rec}})^{1/3}\,,
\end{equation}
where $t_{\rm rec}=1/(\alphaB\nH)$ is the recombination time and
$R_{\rm St,0}= \left[3Q_{\rm i}/(4\pi \alphaB \nH^2)\right]^{1/3}$ is
the dustless Str\"{o}mgren radius \citep{spi78}. For our test, we take
$\nH=10^2 \cm^{-3}$, $Q_{\rm i}=10^{49}\second^{-1}$,
$\sigma_{\rm H} = 6.3 \times 10^{-18} \cm^{2}$, and
$\alphaB = 3.02 \times 10^{-13} \cm^3 \second^{-1}$. The simulation
domain is a cubic box with side of $2.4 R_{\rm st,0}=7.2 \pc$, which
is resolved by a grid of $N_{\rm cell}=128^3$ cells. The optical depth
over one neutral cell is
$\Delta \tau = \nHI \sigma_{\rm H} \Delta x = 109 (128/N_{\rm
  cell}^{1/3}) (Q_{\rm i}/10^{49}\second^{-1})^{1/3} (n_{\rm
  H}/10^2\cm^{-3})^{1/3}$, so that the background medium is highly
optically thick initially. The test simulations are run until
$t=8 t_{\rm rec}$, with the hydrodynamic updates turned off.

\begin{figure}[!t]
  \epsscale{1.2}\plotone{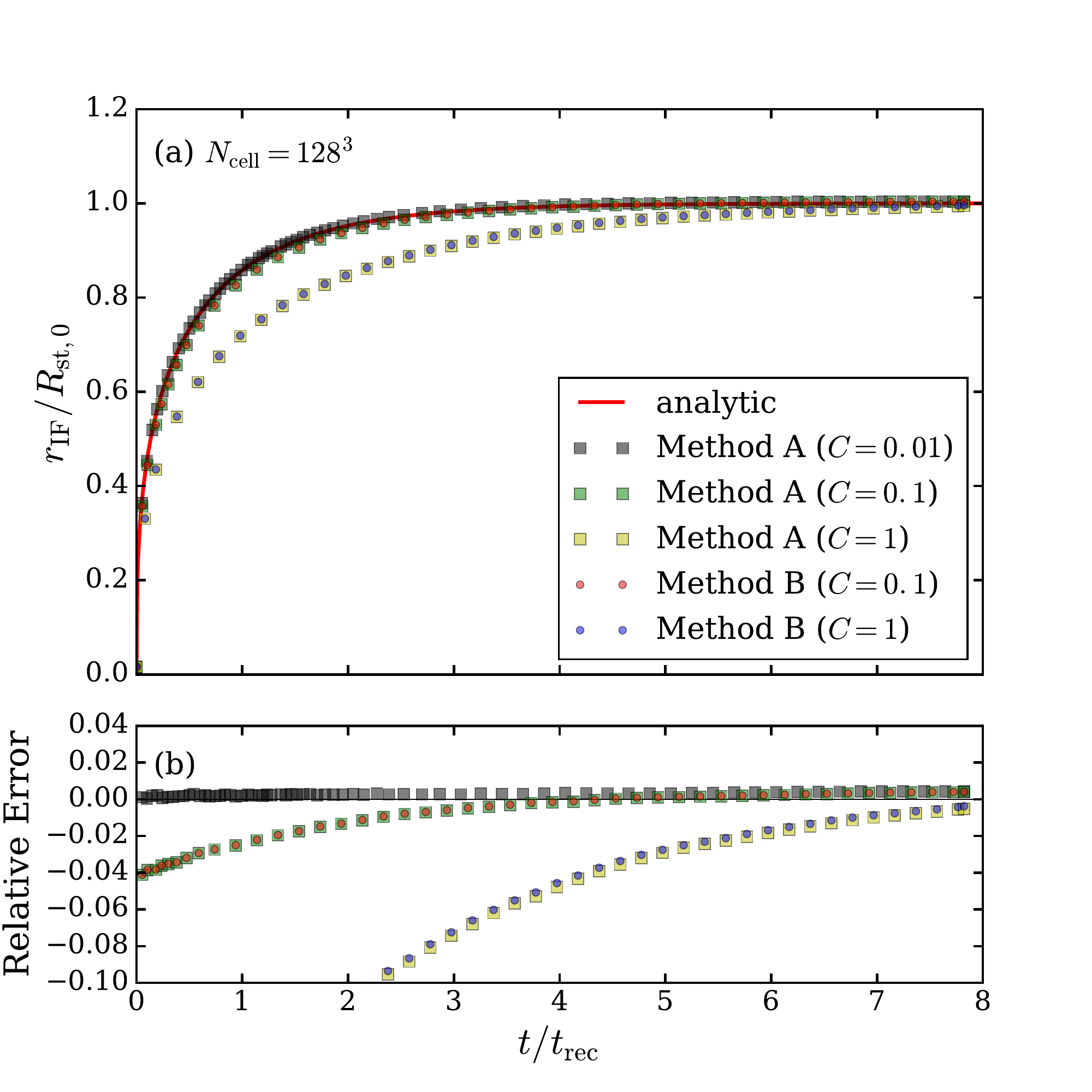}
  \caption{Test of the expansion of an R-type ionization front in a
    uniform medium. Method A (squares) and B (circles) refer to using
    Equations~\eqref{e:xn2} and \eqref{e:xn3}, respectively. (a)
    Ionization front radius {\it vs.} time with different
    time-stepping coefficient $C$ in Equation~\eqref{e:dtss}. The
    analytic solution is shown as the red solid line. (b) Fractional errors
    relative to the analytic solution.}\label{f:IF1}
\end{figure}

\begin{figure}[!t]
  \epsscale{1.2}\plotone{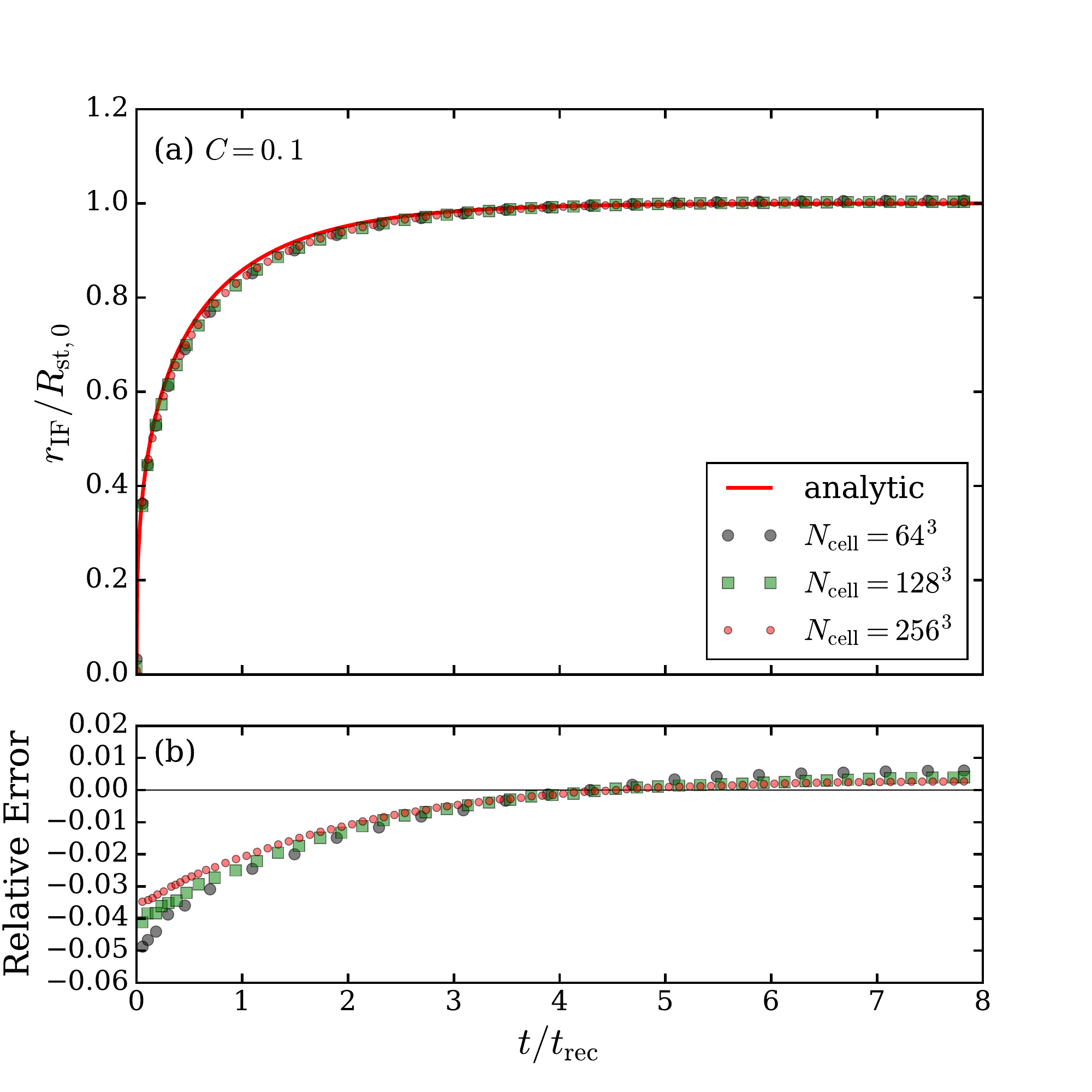}
  \caption{Test of the expansion of an R-type ionization front in a
    uniform medium. Same as \autoref{f:IF1} using Method B with
    $C=0.1$ but with a range of spatial resolution.}\label{f:IF2}
\end{figure}

We first examine the effect of differing time step in the subcycling.
\autoref{f:IF1} plots the temporal changes of $r_{\rm IF}$, defined
as the radius at which $\xn=0.5$, and the relative errors compared to
Equation~\eqref{e:Rtype}. The results based on Equation~\eqref{e:xn2}
(Method A) with $C=0.01$, $0.1$, and $1$ in Equation \eqref{e:dtss}
are shown as squares. For $C=0.1$, the simulation results agree with
the analytic solution within $\lesssim 5\%$, similar to the results of
\citet{bac15}. The results using Equation~\eqref{e:xn3} (Method B)
with $C=0.1$ and $1$ are shown as circles, which are almost identical
to those from Equation~\eqref{e:xn2} with the same $C$. We also
explore the effect of varying spatial resolution by taking the Method
B with $C=0.1$. \autoref{f:IF2} plots the resulting $r_{\rm IF}$
for $N_{\rm cell}=64^3$, $128^3$, and $256^3$, all of which agree
within 2\%. The relative errors are $\sim4$--$5\%$ at early time and
decrease to less than $1\%$ at $t/t_{\rm rec}=8$.

The above results suggest that Method B with $C=0.1$ reproduces the
evolution of R-type ionization fronts quite well. They can be
followed more accurately, albeit at a higher cost, if one restricts
$\Delta t_{\rm ss}$ more strictly, for example, by limiting the
relative changes in $\xn$ to less than 10\% per update
\citep[e.g.,][]{kru07b,mac12}.

\subsubsection{D-type Ionization Front}

As an R-type ionization front approaches $R_{\rm St,0}$, its expansion
speed falls below twice the sound speed $c_{\rm ion}$ in the ionized
region. This develops an isothermal shock in front of the ionization
front, which in turn undergoes a transition to a (weak) D-type front
\citep{shu92}. If the shell of the swept-up material between the
ionization and shock fronts is geometrically thin, the radius of the
shell $r_{\rm sh}$ ought to satisfy the following momentum equation
\begin{equation}\label{e:shell1}
  \dfrac{d}{dt}\left( M_{\rm sh}\dfrac{d r_{\rm sh}}{dt} \right) =
4\pi r_{\rm sh}^2 c_{\rm ion}^2 \rho_{\rm ion}\,,
\end{equation}
where $M_{\rm sh}=(4\pi/3)\rho_0r_{\rm sh}^3$ is the shell mass for
$\rho_0$ the density of the background medium, and
$\rho_{\rm ion} = \rho_0(r_{\rm sh}/R_{\rm St,0})^{-3/2}$ is the
density in the ionized region (assuming instantaneous ionization
equilibrium). \citet{hos06} found that Equation \eqref{e:shell1} has a
self-similar solution
\begin{equation}\label{e:rsh}
  r_{\rm sh} = R_{\rm St,0}\left( 1 + 
    \dfrac{7}{2\sqrt{3}}\dfrac{c_{\rm ion}t}{R_{\rm St,0}}\right)^{4/7}\,.
\end{equation}
\citet{spi78} also solved for a self-similar solution by employing the
requirement of
$\rho_0 (dr_{\rm sh}/dt)^2 \approx c_{\rm ion}^2 \rho_{\rm ion}$ and
derived a similar expression, with $2\sqrt{3}$ in the denominator of
the coefficient in Equation~\eqref{e:rsh} replaced by $4$ (see also
\citealt{bis15}).

\begin{figure}[!t]
  \epsscale{1.2}\plotone{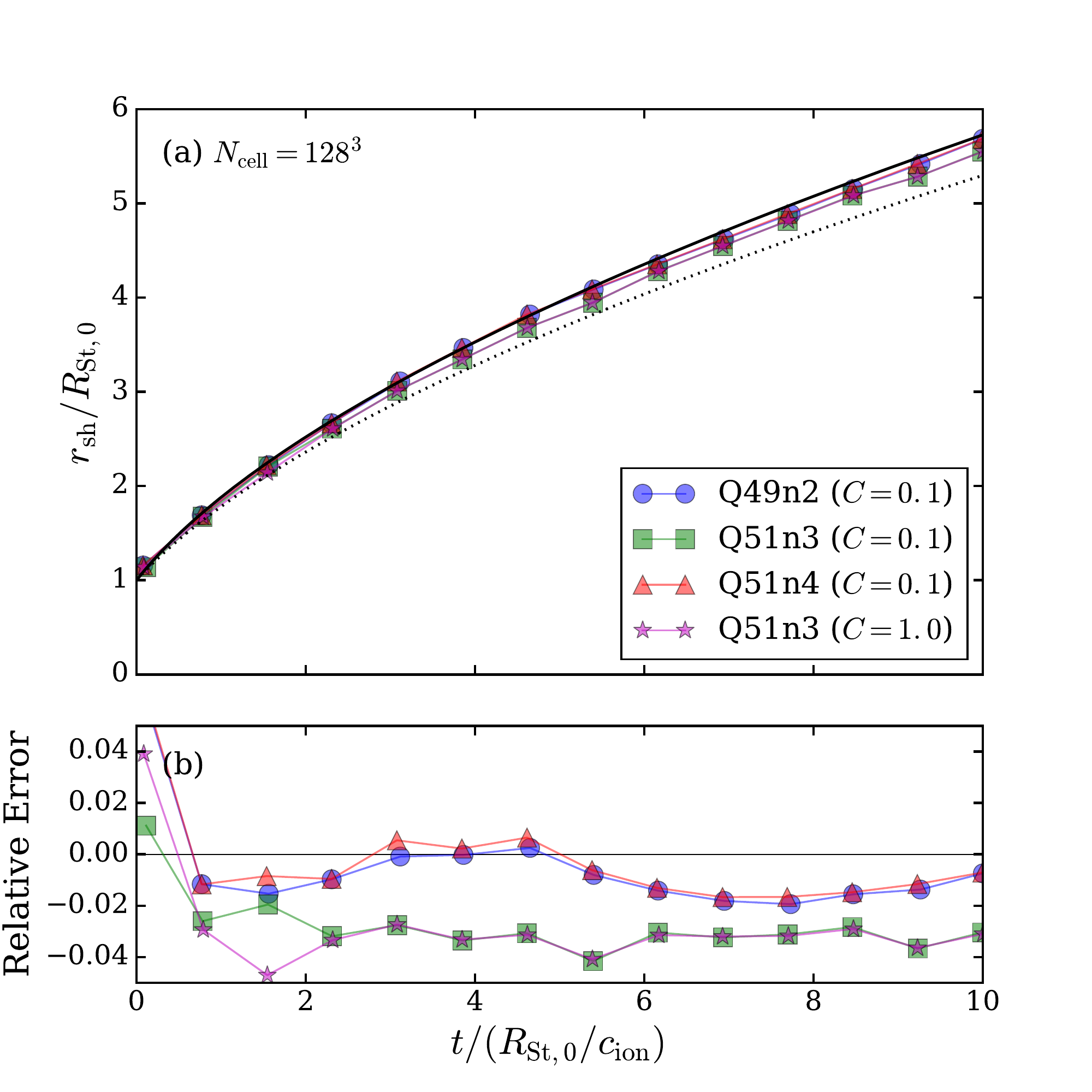}
  \caption{Test of the expansion of a D-type ionization front in a
    uniform medium. (a) Shell radius $r_{\rm sh}$ {\it vs.} time. The
    black solid and dotted lines represent the analytic solution of
    \citet{hos06} and \citet{spi78}, respectively. (b) Errors relative
    to the analytic solution of \citet{hos06}.}\label{f:rsh1}
\end{figure}

For the test of D-type fronts, we set up a cubic domain with $128^3$
cells whose side is $20 R_{\rm St,0}$ long. The domain is filled with
neutral gas with $T_{\rm neu} = 100 \Kel$, and a source placed at the
center starts to emit $Q_{\rm i}$ ionizing photons per unit time from
$t=0$. Unlike in the R-type front tests, we turn on the hydrodynamics
updates so that the gas responds self-consistently to the pressure of
gas at $T_{\rm ion}$ produced by the ionizing radiation. The
simulations are run up to $t=10 R_{\rm St,0}/c_{\rm ion}$ using Method
B with $C=0.1$. At each time, we determine the shell radius
$r_{\rm sh}$ as the position where the gas density is maximal.
\autoref{f:rsh1} plots the resulting $r_{\rm sh}$ and the relative
errors as functions of time for $(Q_{\rm i}/{\rm s}^{-1}$,
$\nH$/cm$^{-3}) = (10^{49}, 10^2)$, $(10^{51}, 10^3)$, and
$(10^{51}, 10^4)$. Shown also as black solid and dotted lines are the
solutions of \citet{hos06} and \citet{spi78}, respectively. Our
results agree with the former better, with typical relative errors
less than 3\%, which is consistent with the results of
\citet{bis15}.\footnote{We note that the benchmark tests in
  \cite{bis15} considered hydrogen-only gas and adopted
  $T_{\rm ion} = 10^4 \Kel$ and
  $\alphaB= 2.7 \times 10^{-13} \cm^3\second^{-1}$, corresponding to a
  $\sim 30\%$ and $\sim 10\%$ difference in $c_{\rm ion}$ and
  $\alphaB$ from ours, respectively. However, our result is unlikely
  to be affected by the specific choice of $T_{\rm ion}$ provided that
  $T_{\rm ion} \gg T_{\rm neu}$.} Using a less stringent time step
size with $C=1.0$, the errors become slightly larger, but the
computational cost of the radiation module is reduced by a factor of
2.5.

\subsubsection{Dusty \ion{H}{2} Region with Radiation Pressure}

In the preceding tests, we have ignored the presence of dust grains
and radiation pressure on them. They are efficient in absorbing (both
ionizing and non-ionizing) UV photons and transfer momentum to the gas
through collisional coupling. The momentum deposition from radiation
pressure may dominate in driving the expansion of \HII\ regions in
dense, massive star-forming environments
\citep{kru09,mur10,fal10,kim16}. Radiation pressure also causes a
non-uniform density distribution inside an \HII\ region in static
equilibrium \citep{dra11}. Time-dependent, spherical models developed
by \citet{kim16} confirmed that when $Q_{\rm i}\nH$ is large, the
expansion of dusty \HII\ regions is dominated by radiation pressure,
with internal structure non-uniform.

Let $L_{\rm n}$ denote the photon luminosity below energy
$h\nuL = 13.6 \eV$ from a source, and $\tau_{\rm edge}$ and
$\rho_{\rm edge}$ be the dust optical depth and gas density at the
edge of the ionized region, respectively. The equation of motion of
the shell is then modified to
\begin{equation}\label{e:shell2}
  \dfrac{d}{dt}\left( M_{\rm sh}\dfrac{d r_{\rm sh}}{dt} \right) =
  \dfrac{L_{\rm n}}{c}e^{-\tau_{\rm edge}} + 4\pi r_{\rm sh}^2 c_{\rm
    ion}^2 \rho_{\rm edge}
\end{equation}
\citep{kim16}. Given $\tau_{\rm edge}$ and $\rho_{\rm edge}$ as a
function of $r_{\rm sh}$ from \citet{dra11}, Equation~\eqref{e:shell2}
can readily be integrated to yield $r_{\rm sh}$ as a function of time,
to which our test results are compared.

We consider an initially uniform medium with $\nH = 10^3 \cm^{-3}$ and
a central source with $Q_{\rm i} = 10^{51} \second^{-1}$ and
$L_{\rm n}=1.5 Q_{\rm i} \hnui$, where $\hnui = 18 \eV$ is the mean
energy of hydrogen ionizing photons. We take a constant dust
absorption cross section
$\sigma_{\rm d} = 10^{-21} \cm^{2}\;{\rm H}^{-1}$ and ionized gas
temperature $T_{\rm ion} = 8000 \Kel$, corresponding to the
  dust opacity parameter of
  $\gamma \equiv (2ck_{\rm B}T_{\rm ion}\sigma_{\rm d})/(\alpha_{\rm
    B} h\nu_{\rm i})=7.58$ (Eq.~7 of
  \citealt{dra11})\footnote{For the numerical value of
    $\gamma$, we took
    $\alpha_{\rm B}\simeq 2.59\times 10^{-13}(T/10^4\;\rm
    K)^{-0.7}\;cm^{3}\;s^{-1}$ from \citet{kru07b}, which is slightly
    different from
    $\alpha_{\rm B}\simeq 2.56\times 10^{-13}(T/10^4\;\rm
    K)^{-0.83}\;cm^{3}\;s^{-1}$ adopted by \citet{dra11}.}. We take a
computational domain with side $20 R_{\rm St,0} = 60 \pc$, resolved by
$128^3$ cells. For comparison, we also run a one-dimensional
simulation in spherical coordinates with the same set of the physical
parameters, but resolving the radial domain $0.1 \pc < r < 30.1 \pc$
using $512$ cells (see \citealt{kim16} for details of simulation
setup).

\begin{figure*}[!t]
  \epsscale{1.0}\plotone{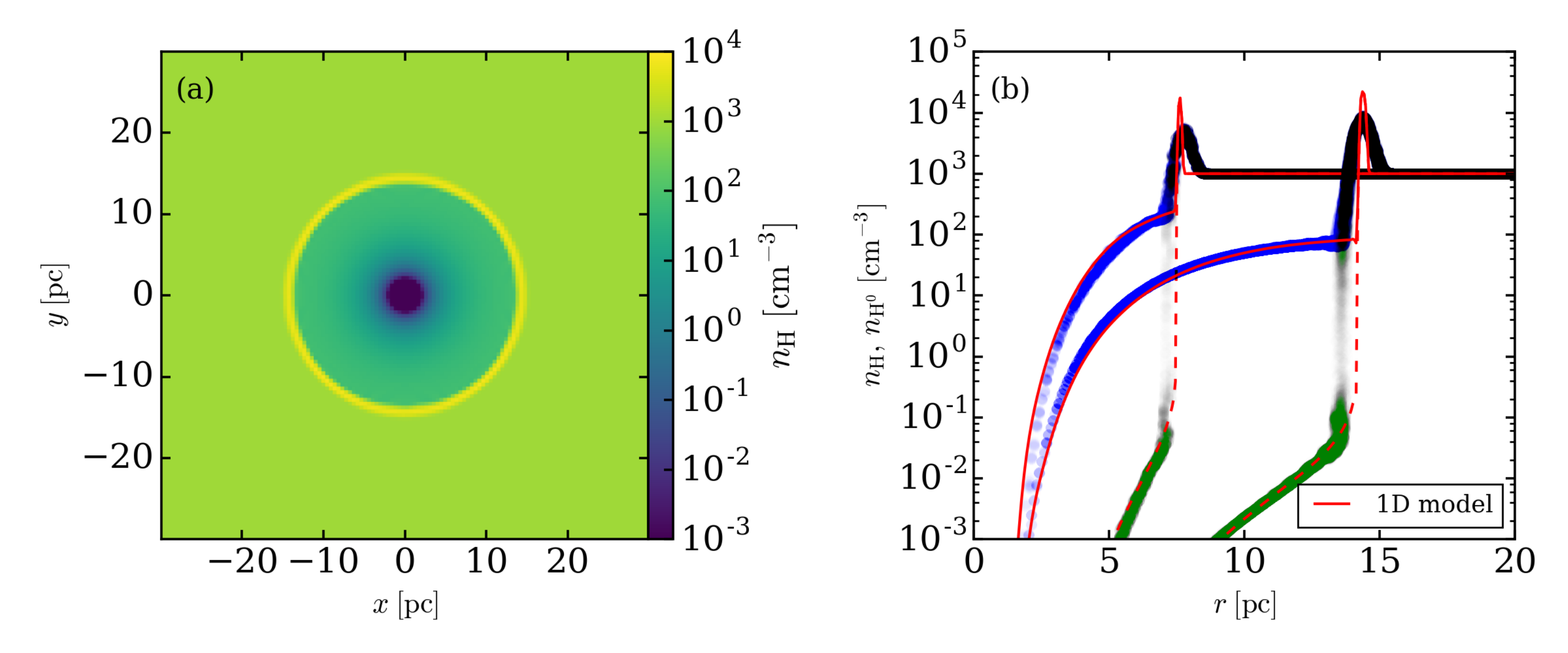}
  \caption{Test of the expansion of a dusty \HII\ region with
    $Q_{\rm i} = 10^{51} \second^{-1}$ and $\nH = 10^3 \cm^{-3}$. (a)
    Slice of $\nH$ through the $z=0$ plane at $t=2.26 \Myr$. (b)
    Radial profiles of $\nH$ (blue) and $\nHI$ (green) at $t=0.67$ and
    $t=2.26 \Myr$. The red lines represent the results of
    one-dimensional simulation in spherical
    coordinates.}\label{f:draine1}
\end{figure*}

\begin{figure}[!t]
  \epsscale{1.2}\plotone{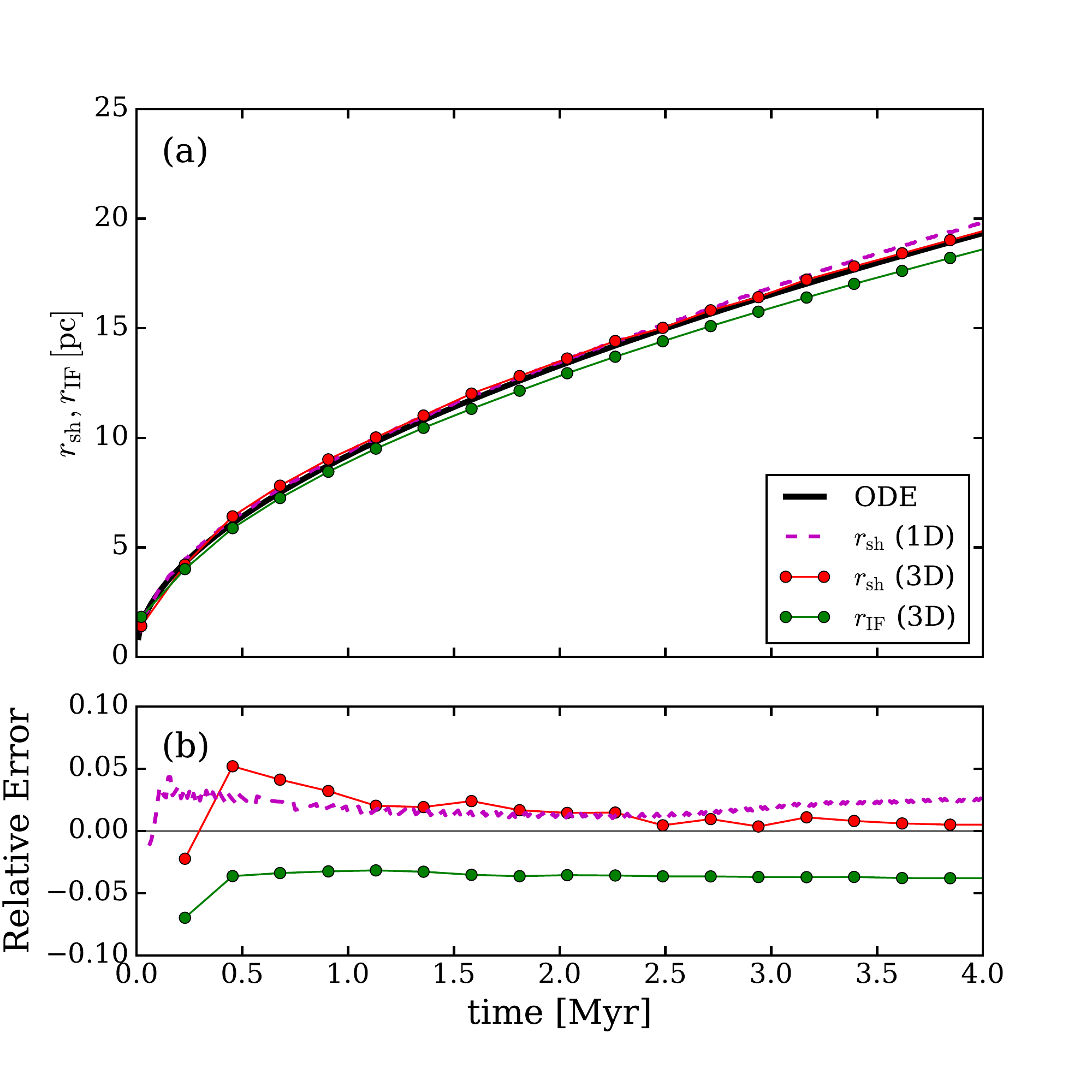}
  \caption{(a) Shell radius (red) and ionization front radius (green)
    {\it vs.}\ time for an expanding dusty \HII\ region with
    $Q_{\rm i} = 10^{51} \second^{-1}$ and $\nH=10^3 \cm^{-3}$. The
    black solid line draws the solution of Equation~\eqref{e:shell2},
    while the dashed line gives the shell radius obtained from the
    one-dimensional simulation. (b) Relative errors with respect to the
    solution of Equation~\eqref{e:shell2}.}\label{f:draine2}
\end{figure}

\autoref{f:draine1} plots (a) the density distribution in the $z=0$
plane at $t=2.26 \Myr$ and (b) the radial profiles of $\nH$ (blue) and
$\nHI$ (green) at $t=0.67$ and $2.26 \Myr$. The corresponding
one-dimensional results are compared as the red lines. As expected,
radiation pressure on dust creates a central cavity devoid of gas and
dust, with the density decaying toward the center approximately as
$\nH \propto \exp(-r_0/r)$, where
$r_0 = \sigma_{\rm d}(Q_{\rm i}\hnui + L_{\rm n})/(8\pi c \kB T_{\rm
  ion})$ \citep{rod16}. Although the shocked shell in the
three-dimensional model has a larger thickness than the
one-dimensional counterpart due to poor spatial resolution, the shell
radius agrees quite well. \autoref{f:draine2} compares
$r_{\rm sh}(t)$ in the simulations with that from the solution of
Equation~\eqref{e:shell2}, again showing good agreement (within 5\%
for $t > 0.5\Myr$) between the numerical and semi-analytic results.
The difference between $r_{\rm IF}$ and $r_{\rm sh}$ in the
three-dimensional model is primarily due to the limited spatial
resolution; the true physical difference between $r_{\rm IF}$ and
$r_{\rm sh}$ is smaller.

\section{APPLICATION TO STAR CLUSTER FORMATION IN TURBULENT CLOUDS}\label{s:comp}

So far our tests have been limited to idealized problems, in which the
background is a uniform medium. By combining our implementation of ART
with other physics modules already existing in the \Athena\ code, we
now demonstrate application of our ART module to an important
practical astronomical problem, namely the formation of a star cluster
in a molecular cloud. We follow the collapse of a turbulent,
self-gravitating cloud and subsequent star formation until the
associated UV radiation feedback from multiple sources halts further
star formation and disperses the remaining gas. In this section, we
first describe the numerical setup and report strong scaling for the
fiducial model. We then compare the non-ionizing UV radiation field
computed by the ART with that based on the $M_1$ closure scheme of
\citet{ski13}, as implemented for point sources in the
single-scattering approximation.

\begin{figure*}[!t]
  \epsscale{0.9}\plotone{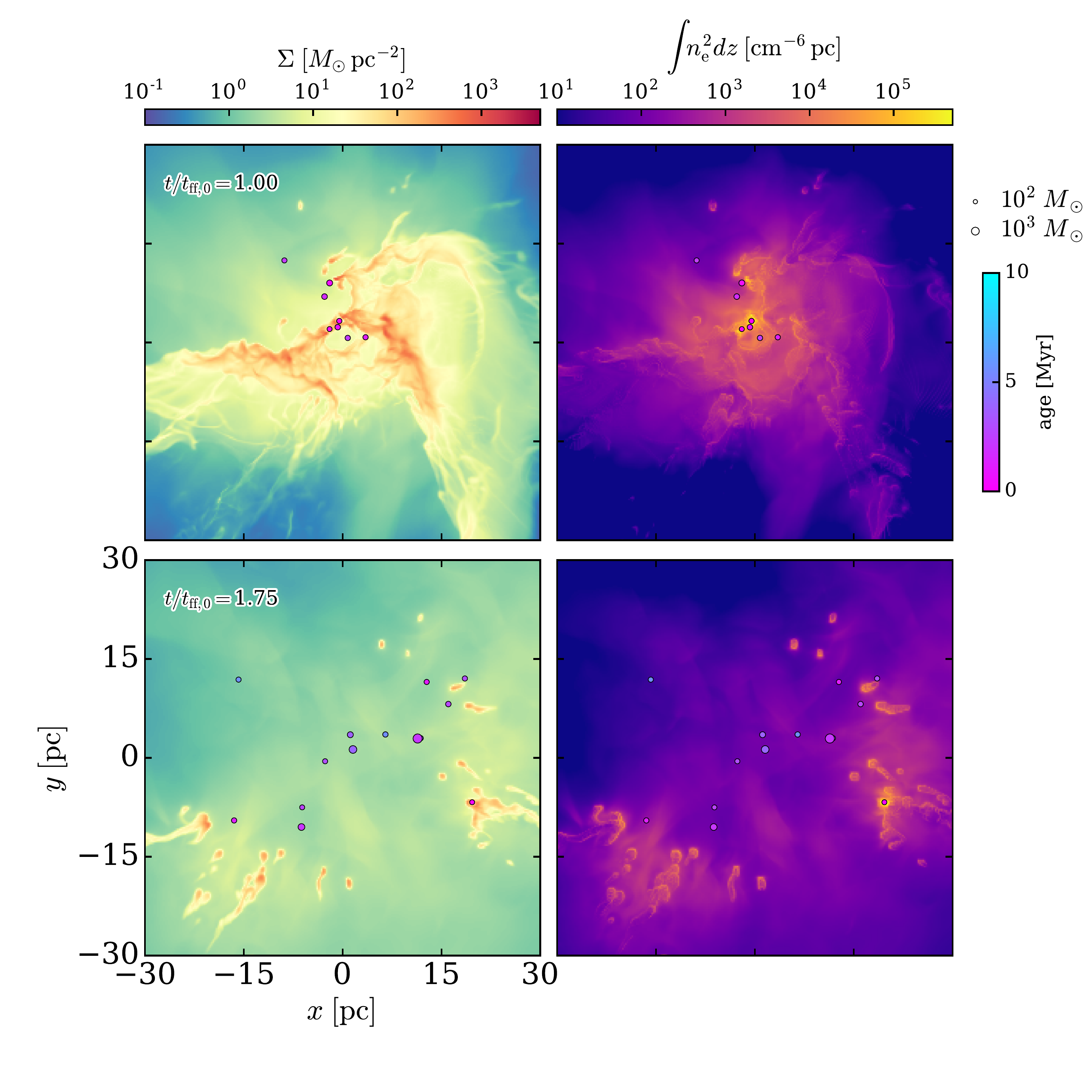}
  \caption{Snapshots of gas surface density (left) and emission
    measure of ionized gas (right) integrated along the $z$-axis at
    $t/t_{\rm ff,0}=1.0$ (top) and $1.75$ (bottom) for the model with
    $M_{\rm cl} = 5\times 10^4\Msun$ and $R_{\rm cl}=15 \pc$. The
    circles mark the projected positions of star particles, with color
    and size representing their age and mass,
    respectively.}\label{f:fiducial}
\end{figure*}

\subsection{Numerical Setup}

We solve Equations~\eqref{e:cont}--\eqref{e:Poisson} to study
evolution of self-gravitating gas interacting with UV radiation from
multiple point sources. We adopt \Athena's HLLC Riemann solver, the
van Leer integrator \citep{sto09}, piecewise-linear spatial
reconstruction, and strict outflow boundary conditions.

We use the sink particle method of \citet{gon13} to treat cluster
formation and gas accretion processes. For this, we create a sink
particle when a cell meets the following three conditions
simultaneously: (1) its density exceeds the threshold corresponding to
the Larson-Penston solution of an isothermal collapse; (2) it is a
local minimum of the gravitational potential; and (3) it has a
converging velocity field along all three principal axes. The mass
accretion rate onto the sink particle is determined by the flux
returned from the Riemann solver at the boundary faces of the $3^3$
ghost cells surrounding the sink particle. The ghost cells in this
sink particle control volume are reset after the hydrodynamics
integration by extrapolation from the nearest active cells. To prevent
spurious mass and momentum flows from the ghost cells to the active
cells, we use the ``diode''-like boundary condition.\footnote{This
  boundary condition is the same as the outflow boundary condition,
  except that the normal velocity component at the boundaries is set
  equal to zero if gas flows from the ghost to active cells.} When two
sink particles come close enough together to make their control
volumes overlap, we simply merge them in such a way that conserves the
total mass and momentum. We have also tested the sink particle method
of \citet{ble14} and found similar results.

The gravitational potential from gas and star particles are calculated
using a FFT Poisson solver with open boundary conditions \citep{ski15}
after mapping star particles' mass onto the mesh via the
triangular-shaped-cloud scheme \citep{gon13}. For the ART, we adopt
$m_{\rm ray} = 4$ and $C=0.1$ in the subcycling and rotate ray
orientation randomly every hydrodynamic cycle. Since the control
volume encompassing a star particle is regarded as a ghost zone, we do
not allow radiation to interact with the gas in the control volume.

Due to limited mass resolution, star particles in our simulation
represent subclusters rather than individual stars. Star particles
emit radiation in two frequency bins: hydrogen ionizing and
non-ionizing photons with luminosity denoted by
$L_{\rm i} = Q_{\rm i} \hnui$ and $L_{\rm n} = L - L_{\rm i}$,
respectively. To assign the luminosity, we first calculate the total
sink mass $M_*$ in the whole domain at a given time. We then use the
stellar population synthesis code SLUG based on the Chabrier initial
mass function \citep{kru15} to calculate the light-to-mass ratio
$\Psi(M_*)$ and the ionizing photon rate per unit mass $\Xi(M_*)$ for
a cluster with mass $M_*$ at birth
\citep{kim16}\footnote{Equation (34) in \citet{kim16} for the
    conversion factor $\Xi(M_*)$ contains typographical errors. The
    correct equation should read
    $\Xi = 10^{46.7 \mathcal{X}^7/(7.28 + \mathcal{X}^7)}
    \second^{-1}\Msun^{-1}$, where
    $\mathcal{X} = \log_{10} (M_*/\Msun)$.}. Finally, a sink particle
with mass $m_*$ is assigned to emit $L=\Psi(M_*) m_*$ and
$Q_{\rm i}=\Xi(M_*) m_*$. Using these conversion factors, for example,
a stellar cluster of mass $(10^2, 10^3, 10^4)\Msun$ has the total
bolometric luminosity and ionizing photon production rate
$(1.1\times 10^4, 7.3\times 10^5, 8.9\times 10^6)\Lsun$ and
$(1.5\times 10^{46},3.5\times 10^{49}, 4.8\times
10^{50})\,\second^{-1}$, respectively. For simplicity, we do not allow
for age variation of $\Psi$ and $\Xi$ and take constant values of
$\hnui = 18 \eV$ for the mean energy of ionizing photons and
$\sigma_{\rm H} = 6.3\times 10^{-18} \cm^{2}$ for the photoionization
cross section.

Our problem initialization is largely similar to the approach of
\citet{ski15} and \citet{ras16}. We consider an isolated,
uniform-density sphere of mass $M_{\rm cl}$ and radius $R_{\rm cl}$
placed at the center of the cubic box with side $L=4R_{\rm cl}$. The
rest of the box is filled with a rarefied medium with density $10^3$
times smaller than that of the cloud. The total gas mass inside the
box is thus $1.014 M_{\rm cl}$. We impose a turbulent velocity field
realized by a Gaussian random distribution with power spectrum
$|\mathbf{v}^2| \propto k^{-4}$ over the wavenumber range
$k \in \left[2,64\right] \times 2\pi/L$ \citep{sto98}. The amplitude
of the velocity field is adjusted such that the total kinetic energy
$E_{\rm kin}$ is equal to the absolute value of the gravitational
potential energy $E_{\rm G} = -\tfrac{3}{5}G M_{\rm cl}^2/R_{\rm cl}$,
making the initial cloud marginally gravitationally bound. The
corresponding virial parameter is
$\alpha_{\rm vir} \equiv 2E_{\rm K}/|E_{\rm G}| = 2$ at $t=0$. For the
particular model presented here, we set
$M_{\rm cl} =5 \times 10^4 \Msun$ and radius $R_{\rm cl}=15\pc$, and
we resolve the simulation box using $256^3$ cells. The initial cloud
conditions are therefore the same as listed in Table 1 for the
fiducial model of \citet{ras16}, although here with
$T_{\rm neu} = 20 \Kel$, the sound speed in the neutral gas is
$c_s=0.27 \kms$ and the opacity for non-ionizing radiation is
$\kappa=500 \cm^2 \gram^{-1}$ (rather than $c_s=0.2\kms$ and
$\kappa=1000 \cm^2 \gram^{-1}$ in \citealt{ras16}), and also, here
$\Psi$ varies with total stellar mass rather than being set to a fixed
value $\Psi = 2000 \erg \second^{-1} \gram^{-1}$.

\subsection{Overall Evolution \& Scaling Test}

\begin{figure}[!t]
  \epsscale{1.2}\plotone{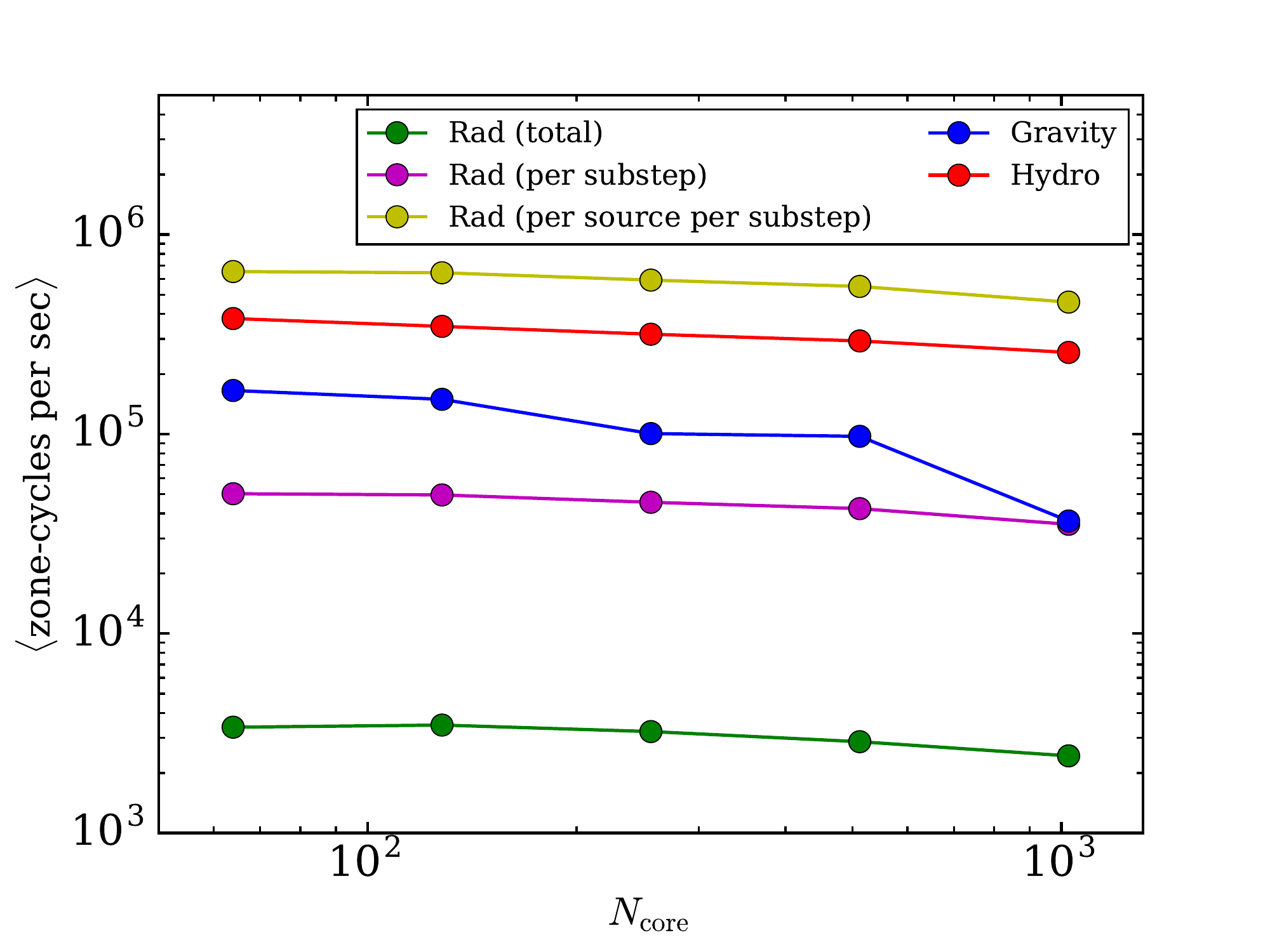}
  \caption{Strong scaling test of a star cluster simulation with
    $N_{\rm cell}=256^{3}$. The average zone-cycles per second during
    $1.16 \leq t/t_{\rm ff,0} \leq 1.19$ against the number of cores
    for different physics modules. During this time interval, there
    are 13 radiation sources and the typical number of substeps taken
    per hydrodynamic update is 14. The parallel efficiency of the
    radiation module from 64 to 1024 cores is 72\%.
  }\label{f:scaling2}
\end{figure}

The initial supersonic turbulence creates filamentary structure via
shock compression. Subsequently, the filaments become gravitationally
unstable and fragment into clumps that undergo runaway collapse to
form stars. The first star particle is spawned at
$t/t_{\rm ff,0}=0.44$, where $t_{\rm ff,0} = 4.3\Myr$ is the initial
free-fall time of the cloud. Star formation continues until
$\sim 1.7 t_{\rm ff,0}$, creating a total of 24 sink particles.
\autoref{f:fiducial} plots the snapshots of gas surface density
(left) and emission measure of the ionized gas (right) on the $x$--$y$
plane at $t/t_{\rm ff,0} = 1.0$ and $1.75$, when 32\% and 99\% of the
final stellar mass has assembled, respectively. The star particle
positions are marked as circles with their size proportional to the
mass. At $t/t_{\rm ff,0} = 2$, the net star formation efficiency is
only 12\%, with the most of the gas ($\gtrsim 70\%$) pushed out of the
simulation box due to photoionization as well as radiation pressure.
We defer to a forthcoming paper a detailed presentation of simulation
results on the star formation efficiency, cloud lifetime, the role of
photoevaporation {\it vs.} radiation pressure in cloud disruption,
etc., and their dependence on the cloud parameters.

We perform a strong scaling test for this realistic star-formation
model, varying the number of cores from $N_{\rm core} = 64$ to 1024.
While the time for the hydrodynamic and self-gravity updates remains
roughly constant throughout the simulation, the cost of the radiation
update scales with the number of sources as well as the number of ART
substeps taken per hydrodynamic update (typically $\sim10$--$20$).
\autoref{f:scaling2} plots the average zone-cycles per second
(i.e., the total number of cells divided by the CPU time) for
different physics modules during the time interval of
$1.16 \le t/t_{\rm ff,0} \le 1.19$ when $\sim 97\%$ of the
computational domain is filled with ionized gas and the number of
point sources is 13. The zone-cycles per second for hydrodynamics and
gravity are weakly-decreasing functions of $N_{\rm core}$ because of
the increasing computation-to-communication ratio. Although the
radiation update is the most expensive part owing to multiple sources
and subcycling, the cost of the ART normalized by the number of
sources and the number of substeps remains roughly constant and is in
fact cheaper than the hydrodynamics update. The relative parallel
efficiency from 64 to 1024 processors is 72\%, 22\%, and 67\% for the
radiation, gravity, and hydrodynamic updates,
respectively.\footnote{Inefficient scaling of the FFT gravity
    module for the largest $N_{\rm core}$ is likely caused by
    communication overhead associated with global transpose of data
    among processors.}

\subsection{Comparison of Radiation Field Computed from $M_1$-Closure
  and ART}

\begin{figure*}[!t]
  \epsscale{1.1}\plotone{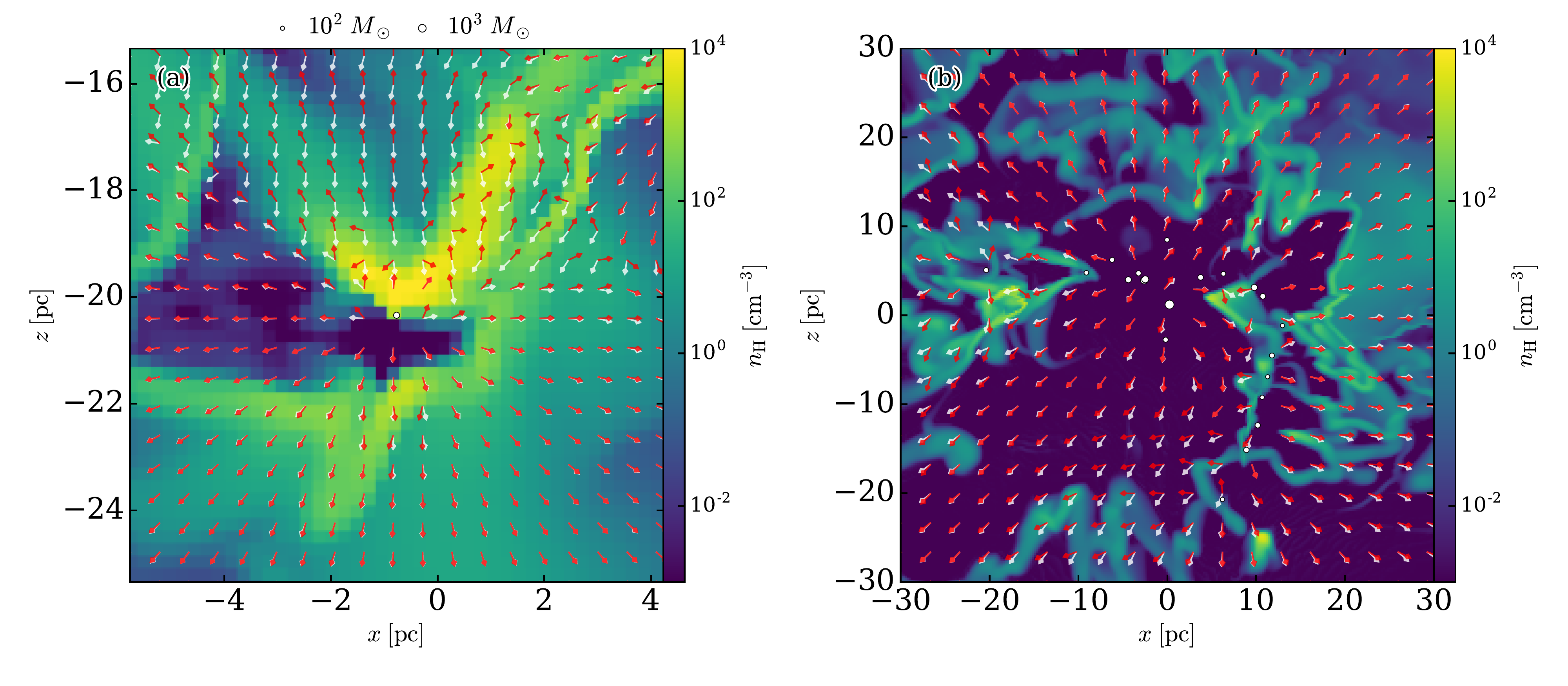}
  \caption{Snapshots of the control model with radiation pressure
    alone in the plane (a) with $y=-0.4\pc$ at
    $t_{\rm 10\%}=0.7 t_{\rm ff,0}$ and (b) with $y=0.3\pc$ at
    $t_{\rm 90\%}=1.75 t_{\rm ff,0}$. All star particles lying within
    $\pm 3\pc$ of the slice plane are shown as white circles. The
    arrows in white and red represent the direction of the radiation
    flux $\Frad$ calculated from the ART and $M_1$ method,
    respectively.}\label{f:comp}
\end{figure*}

\citet{ras16} performed RHD simulations of star formation in turbulent
molecular clouds regulated by non-ionizing radiation pressure on dust
alone. They used the \Hyperion\ code \citep{ski13} employing the
$M_1$-closure relation to evolve the two-moment RT equations. The
\Hyperion\ radiation solver uses explicit integration in time with a
reduced speed of light, employing an HLL-type Riemann solver to
compute radiation fluxes between cells, the piecewise linear spatial
reconstruction, and the first-order backward Euler method for
radiation energy and flux absorption source term updates. For their
fiducial model with $M_{\rm cl} = 5\times 10^4 \Msun$,
$R_{\rm cl} = 15 \pc$, and $\alpha_{\rm vir}=2.0$, \citet{ras16}
obtained a final star formation efficiency of 42\%, larger than
the 12\% found in our ART simulation that considers both
ionizing and non-ionizing radiation, as described above.

To study whether the difference in the star formation efficiency owes
to the effect of ionizing radiation and/or to the method of solving
the RT equations, we have run an additional ART simulation with the
same set of cloud parameters, but by turning off ionizing radiation.
To match the parameters of \citet{ras16}, for this comparison run we
adopt a constant mass-to-light ratio
$\Psi = 2000 \erg \second^{-1}\gram^{-1}$, dust absorption cross
section $\sigma_{\rm d} = 2.34 \cm^2 \gram^{-1}$, and an isothermal
equation of state with isothermal sound speed $c_{\rm s} = 0.2 \kms$.
While overall evolution of this model is very similar to that of
\citet{ras16}, the resulting final star formation efficiency is
$\SFE = 0.25$. This discrepancy in $\SFE$ between the two results can
be largely attributed to the difference in the radiation solvers.

Although the $M_1$-closure scheme has been benchmarked for a variety
of idealized test problems \citep[e.g.,][]{gon07,ski13,ros13,ros15},
it is valuable to check how reliable its radiation field is compared
to that based on the ART method for a problem with complex
distributions of sources and matter. For this purpose, we take the
data at two different epochs from the ART simulation with only
non-ionizing radiation and use them as inputs to the \Hyperion\
radiation solver. With \Hyperion\, we read in the gas density as well
as the positions $\mathbf{r}_{*}$ and masses $m_*$ of all sink
particles, and assign the source function $j_*$ of a Gaussian form to
each sink as
\begin{equation}
  j_*(\mathbf{r}) = \dfrac{m_* \Psi}{(2\pi
  \sigma_*^2)^{3/2}}\exp\left(-\frac{\lvert \mathbf{r} -
    \mathbf{r}_{*}\rvert^2}{2\sigma_*^2}\right)\,,
\label{eq:jsource}
\end{equation}
where the width is taken to be $\sigma_* = (2 \log 2)^{-1/2} \pc$: the
corresponding full width at half maximum is $1 \pc$ \citep{ras16}. We
then perform radiation updates, while fixing hydrodynamic variables
and the star particles. We integrate radiation variables over
$10^3 \yr$, sufficiently long for the radiation field to reach a
steady state.

To explore the differences between the two radiation solutions, we
choose two epochs, $t_{\rm 10\%} = 0.7t_{\rm ff,0}$ and
$t_{\rm 90\%} = 1.75 t_{\rm ff,0}$, when 10\% and 90\% percent of the
final stellar mass has assembled, respectively. The former corresponds
to an early stage of cluster formation when 5 sink particles have been
created, which are mostly embedded in dense nodes of filaments. At
$t=t_{\rm 90\%}$, 39 sources are distributed more evenly around the
center of the domain with half-mass radius $8.2 \pc$; radiation has
cleared out most of the gas in the immediate surroundings. At the
latter time, the system is globally super-Eddington, with the
remaining gas actively expelled from the domain by the radiation
force. The center of mass, $\mathbf{r}_{\rm CM}$, of the sink
particles is $(-1.9, 5.9, -5.6)\pc$ and $(1.1,0.3,0.6)\pc$ at
$t_{\rm 10\%}$ and $t_{\rm 90\%}$, respectively.

\autoref{f:comp} shows the gas structure for the (a) $t_{\rm 10\%}$
and (b) $t_{\rm 90\%}$ epochs, in slices centered at the star particle
located at $\mathbf{r}_{\rm sink}=(-0.8,-0.4,-20.3)\pc$ and the center
of mass of the all star particles
$\mathbf{r}_{\rm CM} = (1.1,0.3,0.6)\pc$, respectively. The white
circles mark the positions of star particles within $3\pc$ from the
slice plane. The arrows indicate the directions of the radiation
fluxes computed by the ART scheme (white) and $M_1$ scheme (red). At
$t_{\rm 10\%}$, there are some discrepancies in the direction of the
radiation fluxes. In the immediate vicinity of the star particle, the
radiation fluxes computed by the ART scheme are directed radially
outward from the sink particle, while those from the $M_1$ scheme have
non-radial component. In the upper middle region, the radiation flux
computed by the ART scheme is directed downward (resulting from a star
particle at $z \sim -15\pc$ outside the panel), while the $M_1$
radiation flux is directed upward. We note that this region
  is heavily shielded by an intervening dense clump located at
  $(x,y) \sim (-1,-20)\pc$ and the radiation field is very weak
  compared to that in the immediate vicinity of the sink or in other
  regions at the same distance from the sink. Therefore, while the
  $M_1$ flux is inaccurate in this region, it is much smaller
  (typically by a factor of 10--30) than the mean value at similar
  distances. At $t_{\rm 90\%}$, the radiation fields from the two
methods are in good overall agreement with each other, although the
flux directions are somewhat different in the lower-middle region
where the gas density is quite low and multiple beams going in
different directions cross.

\begin{figure*}[!t]
  \epsscale{1.0}\plotone{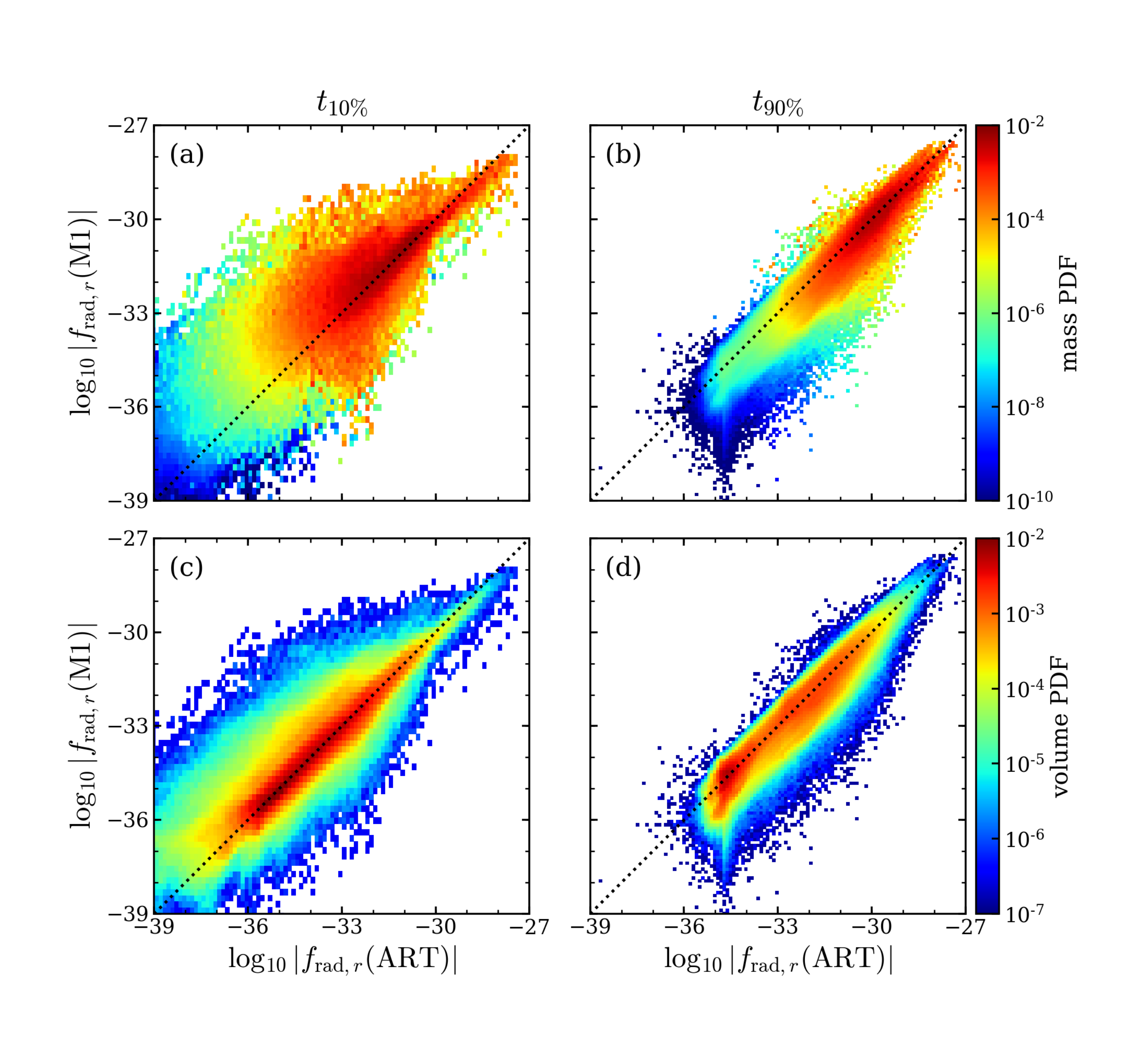}
  \caption{Comparison of radial component (relative
    to the CM) of the radiation force (in units of dyne cm$^{-3}$) computed by the ART and $M_1$
    schemes at times $t_{\rm 10\%}$ (left) and $t_{\rm 90\%}$ (right).
    The upper and lower panels show mass- and volume-weighted
    probability density distributions, respectively. For comparison,
    the dashed lines show the relation that would be obtained if the
    radiation force were identical for the two methods.}
  \label{f:comp2}
\end{figure*}

One quantitative measure of the global dynamical impact of radiation
fields is the radial component, $f_{{\rm rad},r}$, of the radiation
force (per unit volume) relative to $\mathbf{r}_{\rm CM}$. At each
point ${\mathbf r}$ in the domain, this is given by
\begin{equation}\label{eq:fradr}
  f_{{\rm rad},r} = \dfrac{\chi \Frad}{c}\cdot
  \dfrac{\mathbf{r} - \mathbf{r}_{0}}{|\mathbf{r} - \mathbf{r}_{0}|}
  \,,
\end{equation}
where $\mathbf{r}_0=\mathbf{r}_{\rm CM}$. \autoref{f:comp2}
compares the probability distribution functions (pdfs) of
$|f_{{\rm rad},r}|$ computed from ART and from the $M_1$ method at
$t=t_{\rm 10\%}$ (left) and $t_{\rm 90\%}$ (right); pdfs are shown
weighted by the cell mass (top) and volume (bottom). Overall, the
$M_1$ method reproduces the ART radiation field reasonably well,
with $|f_{{\rm rad},r}\, (M_1)|/|f_{{\rm rad},r}(\rm ART)|\approx 1.0$ at $t_{\rm 10\%}$ and 0.89 at $t_{\rm 90\%}$, when averaged over the entire range of $|f_{{\rm rad},r}|$.
At $t=t_{\rm 10\%}$ there is a significant scatter relative
to the one-to-one relationship (dashed lines), which is likely to be caused by the inability of the $M_1$ scheme to describe the superposition of streaming
radiation from multiple sources going in different directions. The
scatter is small at $t=t_{\rm 90\%}$ because the radiation sources are
more centrally concentrated relative to the gas.

A careful examination of \autoref{f:comp2}(a) shows that
$|f_{{\rm rad},r}\, (M_1)|/|f_{{\rm rad},r}(\rm ART)|\approx 0.55$ for the largest values of $|f_{{\rm rad},r}(\rm ART)| \gtrsim 10^{-29}\rm\;dyne\;cm^{-3}$  at $t=t_{\rm 10\%}$. The corresponding physical locations are in
immediate regions surrounding sink particles where both density and
radiation flux are high. This discrepancy is suggestive of a possible
reason why the total star formation efficiency in the the \Hyperion\
simulation is larger: with a lower radiation force in the immediate
vicinity of sink particles, accretion of material onto sinks may not
be limited as strongly as it is in the ART simulation.

\begin{figure}[!t]
  \epsscale{1.2}\plotone{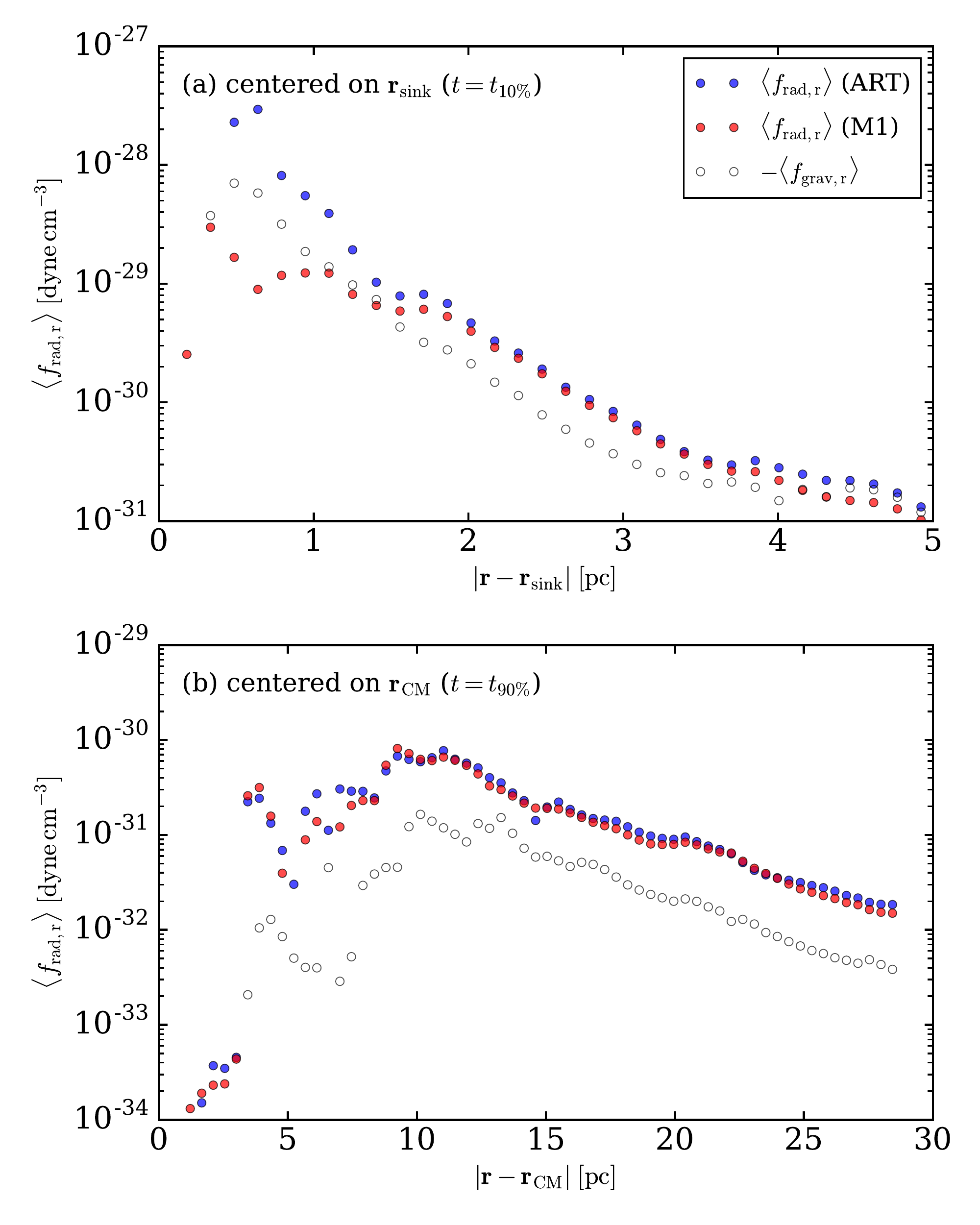}
  \caption{Angle-averaged radial radiation force computed by ART
    (blue) and $M_1$ (red) centered (a) at a sink particle at
    $t=t_{\rm 10\%}$, and (b) at the center of mass of all sinks at
    $t=t_{\rm 90\%}$. The absolute values of the angle-averaged
    gravitational forces are shown as open circles. }\label{f:comp3}
\end{figure}

To investigate this issue in more detail, we examine the radial
component of the radiation force centered on individual star
particles. This is computed as in Equation (\ref{eq:fradr}), except
now $\mathbf{r}_0$ is the location $\mathbf{r}_{\rm sink}$ of an
individual star particle. For example, \autoref{f:comp3}(a)
compares the angle-averaged radiation force
$\langle f_{{\rm rad},r} \rangle = \int f_{{\rm rad},r} d\Omega/\int
d\Omega$ relative to a sink particle located at
$\mathbf{r}_{\rm sink}=(-0.8,-0.4,-20.3) \pc$, isolated from other
sink particles. Although the $M_1$ radiation force is in good
agreement with the ART radiation force for
$|\mathbf{r} - \mathbf{r}_{\rm sink}| > 1.5\pc$, the former
significantly underestimates the radiation force within
$\lesssim 1.5\pc$ of the source. Since the source function in the
$M_1$-scheme (Equation [\ref{eq:jsource}]) is smoothed over the finite
width of $\sim 1\pc$, it takes $\sim4$--$5$ cells to build up the flux
expected for a point source radiation. In contrast, the ART scheme
with the adopted parameters has been proven successful in reproducing
a point source radiation field with errors of only a few percent (see
Section~\ref{s:vacuum}). Interestingly, the magnitude of the
gravitational force (open circles) at
$|\mathbf{r} - \mathbf{r}_{\rm sink} | \lesssim 1\pc$ is larger than
the radiation force from the $M_1$ scheme, but smaller than that from
the ART scheme. This suggests that the direct radiation pressure
feedback in the former is less effective in halting accretion than in
the latter. The difficulty in resolving flux $\propto r^{-2}$ near
point sources in the $M_1$ scheme may explain the difference in the
star formation efficiency between the two simulations.

Although the radiation fields from the two methods are quite different
within a few zones of the point sources, they agree at both larger
distances from individual sources and on global scales, particularly
at late times. \autoref{f:comp3}(b) plots the angle-averaged radial
force as computed with ART and with $M_1$ as functions of the distance
from the center of mass of all sinks at $t=t_{\rm 90\%}$. Despite a
complex source geometry, the results from the two methods agree very
well. The fraction of radiation escaping the surface of a sphere of
radius $r_{\rm max}=2R_{\rm cl} - |\mathbf{r}_{\rm CM}|=29 \pc$
centered at $\mathbf{r}_{\rm CM}$ is equal at $f_{\rm esc}=0.63$
between the two methods.

We also calculate the normalized volume-integrated radial force
\begin{equation}\label{e:frad_tot}
  \mathcal{F}_{\rm out} = \frac{4\pi c}{(1 - f_{\rm esc})L_{*,{\rm
        tot}}} \int_0^{r_{\rm max}} \langle f_{{\rm rad},r} \rangle
  r^2 dr\,,
\end{equation}
which would be equal to unity had all sources with total luminosity
$L_{*,{\rm tot}}$ been located at $\mathbf{r}_{\rm CM}$. For multiple
sources distributed in space, $\mathcal{F}_{\rm out}$ is reduced due
both to flux cancellation and to misalignment of the radiation force
vectors from the radial direction toward $\mathbf{r}_{\rm CM}$. The
values of $f_{\rm esc}$ and $\mathcal{F}_{\rm out}$ from the ART and
$M_1$ schemes are quite comparable to each other. Table~\ref{table}
summarizes these global properties of the radiation fields for the two
radiation solvers at $t=t_{\rm 10\%}$, $t_{\rm 50\%}$, and
$t_{\rm 90\%}$.

\capstartfalse
\begin{deluxetable}{cccccc}
\tabletypesize{\footnotesize}
\tablewidth{0pt}
\tablecaption{Global properties of the
radiation fields from the ART and $M_1$ schemes\label{table}}\tablehead{
\colhead{$f_*$ (\%)} & %
\colhead{$t/t_{\rm ff,0}$} & %
\colhead{$f_{\rm esc,ART}$} & %
\colhead{$f_{\rm esc,M1}$} & %
\colhead{$\mathcal{F}_{\rm out, ART}$} & %
\colhead{$\mathcal{F}_{\rm out, M1}$}   \\
\colhead{(1)}  & \colhead{(2)} & \colhead{(3)} & \colhead{(4)} &
\colhead{(5)}  & \colhead{(6)}}
\startdata
10 & 0.7  & 0.29 & 0.28 & 0.079 & 0.073  \\
50 & 1.0  & 0.29 & 0.30 & 0.20 & 0.15  \\
90 & 1.75 & 0.63 & 0.63 & 0.64 & 0.59 \\
\enddata
\tablecomments{Column (1) is the fraction of the total stellar mass
  formed. Column (2) is the time in units of the initial free-fall
  time. Columns (3) and (4) are the escape fraction of the photons
  through an enclosing sphere computed by the ART and $M_1$ methods,
  respectively. Columns (5) and (6) are the normalized,
  volume-integrated radial force from the ART and $M_1$ methods,
  respectively. The measurements of $f_{\rm esc}$ and
  $\mathcal{F}_{\rm out}$ are made for spheres with radii $22 \pc$,
  $25 \pc$, and $29\pc$ centered at $\mathbf{r}_{\rm CM}$ at
  $t_{\rm 10\%}$, $t_{\rm 50\%}$, and $t_{\rm 90\%}$, respectively.}
\end{deluxetable}
\capstarttrue

\section{SUMMARY}\label{s:summary}

Radiation feedback from young massive stars has profound influence on
the evolution of their natal clouds and subsequent star formation.
Stellar radiation that escapes from dense molecular clouds is also
essential in controlling the thermal and ionization states of the
diffuse ISM in galaxies and contributes to ionization of the IGM at
even further distances. Because of the long-range, multiscale nature
of the interactions between matter and radiation in three dimensions
and of the highly inhomogeneous spatial structure of the gas and
radiation source distribution, it is non-trivial to handle RT properly
in hydrodynamic simulations. The ART algorithm developed by
\citet{abe02} provides an accurate means of treating RT from point
sources, which retains spatial resolution at moderate cost by
splitting rays as the distance from a given source increases.
\citet{ros17} recently improved the parallel performance of the ART
method by implementing completely non-blocking, asynchronous MPI
communication.

In this paper, we describe our implementation of ART in the Eulerian
grid-based code \Athena\ and present results of performance tests. We
adopt the non-blocking, asynchronous parallelization algorithm
suggested by \citet{ros17} for exchanges of information along rays
between processors. We further improve the parallel performance by (1)
passing photon packets to neighbor processors whenever a certain
number of local ray-traces have been completed and (2) making use of
one-sided communications to update the ``destroy count'' of terminated
rays. The radiation source terms, hydrogen photoionization, radiation
pressure, etc.\ are all substepped (relative to the hydrodynamic
timestep) and updated in an operator-split manner after each ART
sweep, with the substepping time interval set by the
ionization/recombination time via Equation (\ref{e:dtss}). To update
the hydrogen ionization state we use an approximate solution
(Equation~[\ref{e:xn3}]) of the rate equation, which we find is more
numerically robust and gives results that are essentially the same as
the full analytic solution (Equation~[\ref{e:xn2}]).

We have verified the performance and accuracy of our implementation of
the ART scheme on a wide variety of test problems. The results of weak
and strong scaling tests (Figure \ref{f:scaling}) show that the cost
of the ART (per source) remains comparable to that of the hydrodynamic
update on up to $10^3$ processors. The vacuum radiation test shows
that if the number of rays passing through a cell is $m_{\rm ray}=4$,
the median value of the errors in the calculated radiation energy
density is only $\sim1$--$4\%$, and the accuracy can be further
improved by rotating the ray directions randomly \citep{kru07b} at
each step (\autoref{f:vacuum}). Through standard test problems for
the expansion of \HII\ regions, we demonstrate that our radiation
solver reproduces quite well the expected solutions of expanding
R-type and D-type ionization fronts as well as the expansion of a
dusty \HII\ region with radiation pressure
(Figures~\ref{f:IF1}--\ref{f:draine2}).

As a practical application demonstrating the use of our code, we
conduct a simulation of star cluster formation in a turbulent,
self-gravitating molecular cloud with
$M_{\rm cl} = 5\times 10^4 \Msun$, $R_{\rm cl} = 15 \pc$, and initial
virial parameter of $\alpha_{\rm vir}=2$. We find that the net star
formation efficiency is $\SFE = 0.12$, with most of the gas expelled
from the simulation box when both photoionization and radiation
pressure from UV radiation are present. The strong scaling test for
this problem (Figure \ref{f:scaling2}) shows that the parallel
efficiency of the ART module is as good as that of the hydrodynamics
module. The total cost is however dominated by the radiation update,
which involves multiple sources as well as subcycling.

We have also run an analogous model with radiation pressure alone
(i.e., without photoionizing radiation), in order to directly compare
with results obtained from the \Hyperion\ code, which solves the
two-moment radiation equations with an $M_1$ closure relation.
Considering an identical radiation source and density distribution,
the radiation field computed using the $M_1$ scheme agrees with that
from ART on large scales, even for distributed sources. Since,
however, point sources are smoothed over a finite volume in \Hyperion,
it is not able to reproduce the radiation field accurately near
individual sources. As a consequence of the reduced radiation flux
near point sources, radiation feedback is less able to limit accretion
of nearby material, which likely accounts for the increased net star
formation efficiency found with \Hyperion\ ($\SFE=0.42$) compared to
that with our ART implementation ($\SFE=0.25$). We conclude that one
should be cautious when modeling point sources using the $M_1$ scheme
if radiation feedback is important to limiting accretion; one approach
might be to expand the control volume of each sink particle such that
the radiation flux is well resolved at its boundary.

The test results presented in this paper confirm that our
implementation of ART in the \Athena\ code is accurate and efficient.
In a subsequent work, we shall present results from application of the
code to cluster formation and radiation feedback in turbulent
molecular clouds. With an accurate and efficient method for treating
the effects of radiation, we are able to survey a range of parameters,
studying the dependence on the cloud mass and surface density of the
star formation efficiency and cloud lifetime, as well as the relative
roles of photoevaporation and radiation pressure in shaping and
disrupting GMCs.

\acknowledgements The authors thank Anna Rosen for making a copy of
her code and manuscript on adaptive ray tracing available to us before
publication. J.-G.K. wishes to thank Shane Davis, Chang-Goo Kim, Jim
Stone, Kengo Tomida, and G\'{a}bor T\'{o}th for their helpful
discussions and advice. J.-G.K. acknowledges support from the National
Research Foundation of Korea (NRF) through the grant
NRF-2014-Fostering Core Leaders of the Future Basic Science Program.
The work of W.-T.K.\ was supported by the National Research Foundation
of Korea (NRF) grant funded by the Korea government (MEST) (No.\
3348-20160021). The work of E.C.O.\ on this project was supported by
the U.S. National Science Foundation under grant AST-1312006, and by
NASA under grant NNX14AB49G. This work was performed under the
auspices of the U.S. Department of Energy by Lawrence Livermore
National Laboratory under Contract DE-AC52-07NA27344. The computation
of this work was supported by the Supercomputing Center/Korea
Institute of Science and Technology Information with supercomputing
resources including technical support (KSC-2015-C3-049), and the
PICSciE TIGRESS High Performance Computing Center at Princeton
University.
\\
\textit{Software:} Athena \citep{sto08}, SLUG \citep{kru15}, yt \citep{tur11}, numpy
\citep{van11}, matplotlib \citep{hun07}, IPython \citep{per07}, pandas
\citep{mck10}.

\appendix

\begin{figure}
  \epsscale{1.2}\plotone{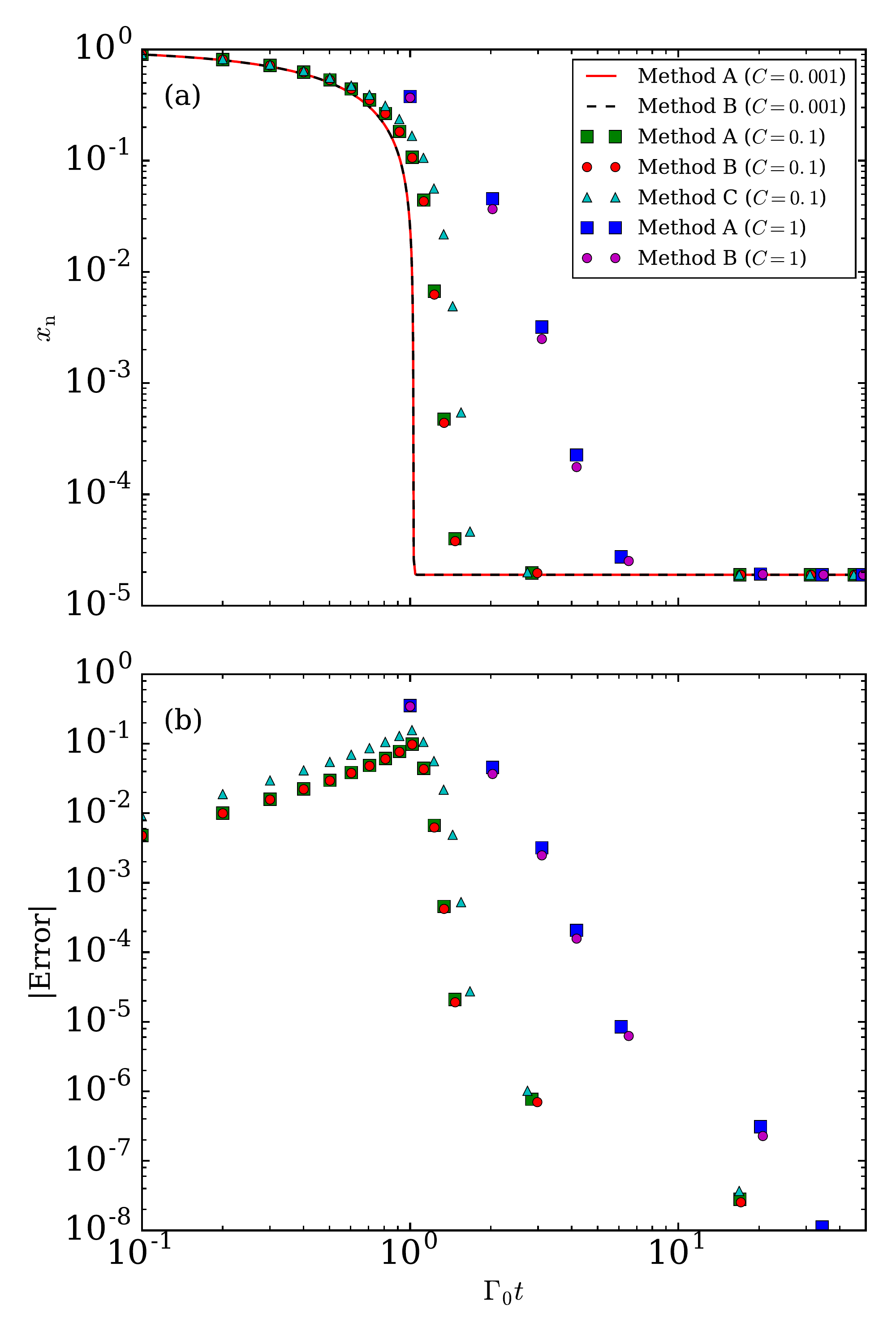}
  \caption{(a) Neutral fraction $\xn$ {\it vs.} time obtained using the
    different update schemes for $\xn$ with various $C$. Methods A and
    B use Equations~\eqref{e:xn} and \eqref{e:xn2}, respectively,
    while Method C employs Equation~\eqref{e:bdf1}. (b) Difference of
    $\xn$ relative to the results with $C=0.001$, shown as the red
    solid line in (a). }\label{f:one-zone}
\end{figure}

\section{One-Cell Test of Photoionization Update}

Here we present test results for the hydrogen photoionization update.
We consider a single cell with width $\Delta x=2\pc$ initially filled
with completely neutral hydrogen of density $\nH$. The cell is exposed
to a fixed ionizing radiation field $Q_{\rm i}=10^{49} \second^{-1}$
from $t=0$. The temporal evolution of the neutral hydrogen fraction in
the cell is described by Equation~\eqref{e:xn}, with the
photoionization rate given as
\begin{equation}
  \Gamma = \dfrac{1}{\nH\xn}\dfrac{Q_{\rm i}}{(\Delta x)^3}\left( 1
    - e^{-\nH\xn\sigma_{\rm H}\Delta x} \right)\,.
\end{equation}
The cell has initial optical depth
$\tau_0=\nH\sigma_{\rm H}\Delta x = 3880 \gg 1$, and initial ionization rate
$\Gamma_0 = Q_{\rm i}/(\nH \Delta x^3) = 1.7 \times
10^{-6}\second^{-1} \approx 14 \alphaB\nH$. The cell is expected to be
ionized on the time scale $\Gamma_0^{-1}$, eventually
settling into a balanced state with equilibrium neutral fraction
of $\xneq \approx \alphaB\nH/(\Gamma_0\tau_0) = 1.8 \times 10^{-5}$.

To evolve $\xn$ according to Equation~\eqref{e:xn}, we try three
different schemes. First, we directly use the exact expression of
Equation~\eqref{e:xn2} with varying values of $C$ in
Equation~\eqref{e:dtss}, which we call Method A. The second method
(Method B) uses Equation~\eqref{e:xn3}, again with varying $C$. We
also use a semi-implicit difference method (Method C) to discretize
Equation~\eqref{e:xn} as
\begin{equation}
  \dfrac{\xn^{n+1} - \xn^{n}}{\Delta t_{\rm ss}} = \alphaB\nH(1 -
  \xn^n)(1 - \xn^n) - \xn^{n+1}\Gamma\,,
\end{equation}
which gives a recurrence relation
\begin{equation}\label{e:bdf1}
  \xn^{n+1} = \dfrac{\xn^n + (1-\xn^n)^2\alphaB\nH\Delta t_{\rm ss}}{1
    + \Gamma\Delta t_{\rm ss}}\,,
\end{equation}
where the superscripts ``$n$'' and ``$n+1$'' denote the values at
$t=t_0$ and $t=t_0 + \Delta t_{\rm ss}$, respectively.

\autoref{f:one-zone} plots the resulting temporal changes of $\xn$
from the various methods with different $C$. The results of Methods A
and B with $C=0.001$, which are almost identical to each other, can be
regarded as the true solution. The neutral fraction settles to an
equilibrium value at $t \Gamma_0 \gtrsim 1$. The bottom panel plots
the errors relative to the $C=0.001$ results. It is remarkable that
the difference between the results of Methods A and B is very small
even for $C$ as large as unity and that they are better than those
from Method C. The error is largest when $\xn \sim 0.1$--$0.2$, for
which $|d\xn/dt|$ is quite large.

\bibliography{myref}

\end{document}